\documentclass[10pt,letterpaper]{article}
\usepackage[top=0.85in,footskip=0.75in]{geometry}
\usepackage{amsmath,amssymb}
\usepackage{changepage}
\usepackage[utf8x]{inputenc}
\usepackage{textcomp,marvosym}
\usepackage{cite}
\usepackage{nameref,hyperref}

\usepackage{microtype}
\DisableLigatures[f]{encoding = *, family = * }

\usepackage[table]{xcolor}

\usepackage{array}

\newcolumntype{+}{!{\vrule width 2pt}}

\newlength\savedwidth

\usepackage{setspace} 

\raggedright
\usepackage[aboveskip=1pt,labelfont=bf,labelsep=period,justification=raggedright,singlelinecheck=off]{caption}

\makeatletter
\renewcommand{\@biblabel}[1]{\quad#1.}
\makeatother

\date{}

\usepackage{lastpage,fancyhdr,graphicx}
\usepackage{epstopdf}
  
\newcommand\remove[1]{}
\usepackage{amsfonts}
\usepackage{caption,subcaption}
\usepackage{url}
\usepackage{longtable}
\usepackage{pdflscape}
\usepackage{enumitem}
\usepackage{float}

\usepackage{etoolbox}
\preto\longtable{\par\singlespacing}
\preto\tabular{\par\singlespacing}

\definecolor{darkgreen}{RGB}{0,100,0}

\makeatletter
\newenvironment{chapquote}[2][2em]
  {\setlength{\@tempdima}{#1}%
   \def\chapquote@author{#2}%
   \parshape 1 \@tempdima \dimexpr\textwidth-2\@tempdima\relax%
   \itshape}
  {\par\normalfont\hfill--\ \chapquote@author\hspace*{\@tempdima}\par\bigskip}
\makeatother

\begin{document}
\vspace*{0.2in}

\begin{flushleft}
{\Large
\textbf\newline{
Using Social Media To Predict the Future: A Systematic Literature Review} 
}
\newline
\\
Lawrence Phillips\textsuperscript{1*, \Yinyang},
Chase Dowling\textsuperscript{1, 2, \Yinyang},
Kyle Shaffer\textsuperscript{1},
Nathan Hodas\textsuperscript{1},
Svitlana Volkova\textsuperscript{1}
\\
\bigskip
\textbf{1} Data Sciences and Analytics Group, Pacific Northwest National Laboratory, Richland, Washington, United States of America \\
\textbf{2} Electrical Engineering, University of Washington, Seattle, Washington, United States of America
\\
\bigskip

%
%
\Yinyang These authors contributed equally to this work.





* Lawrence.Phillips@pnnl.gov

\end{flushleft}


\section*{Abstract}
Social media (SM) data provides a vast record of humanity's everyday thoughts, feelings, and actions at a resolution previously unimaginable. Because user behavior on SM is a reflection of events in the real world, researchers have realized they can use SM in order to forecast, making predictions about the future. The advantage of SM data is its relative ease of acquisition, large quantity, and ability to capture socially relevant information, which may be difficult to gather from other data sources. Promising results exist across a wide variety of domains, but one will find little consensus regarding best practices in either methodology or evaluation. In this systematic review, we examine relevant literature over the past decade, tabulate mixed results across a number of scientific disciplines, and identify common pitfalls and best practices. We find that SM forecasting is limited by data biases, noisy data, lack of generalizable results, a lack of domain-specific theory, and underlying complexity in many prediction tasks. But despite these shortcomings, recurring findings and promising results continue to galvanize researchers and demand continued investigation. Based on the existing literature, we identify research practices which lead to success, citing specific examples in each case and making recommendations for best practices. These recommendations will help researchers take advantage of the exciting possibilities offered by SM platforms.


\section*{Introduction}

\begin{chapquote}{Yogi Berra}
``Forecasting is a difficult business, particularly when it is about the future.''
\end{chapquote}

Now more than ever before, companies, governments, and researchers can gather and access data about people on a massive scale. Putting a finger on the pulse of public opinion is made increasingly possible thanks to the rise of social media (SM; for a more comprehensive review of SM platforms see~\cite{batrinca2015social, kwak2010twitter}). SM are Internet-enabled platforms that provide users with a persistent online identity and means of sharing information with friends, families, coworkers, and other users. 
There are many different SM platforms, each of which targets a different aspect of what users want or need: e.g., LinkedIn targets professional networking activities, Facebook provides a means of connecting friends and family, and Twitter provides a platform from which to quickly broadcast thoughts and ideas. These platforms are incredibly popular: as of June 2016, Facebook sees an average of 1.13 billion daily users, including nearly half the populations of the United States~\cite{uscb} and Canada~\cite{statcan} logging in every day \cite{fbstats}. 

Being so widely used, SM platforms generate massive quantities of data. According to~\cite{twitter2013new}, in 2013 users were posting an average of over 500 \emph{million} tweets every day. 
While traditional data sources and records of daily human activity, such as newspapers and broadcast media, are often constrained by national, cultural, and linguistic boundaries, SM platforms are generally consistent provided a user has access to the Internet. Moreover, traditional media requires time to compile relevant information for publication, while SM data is generated in real time as events take place.  

All of this information can be collected and mined by virtually anyone who wishes to use it. As far back as 2009, the United States Geological Survey (USGS) began investigating the possibility of using SM data to detect earthquakes in real time~\cite{twitterEarthquake}. Information about an earthquake spreads faster on SM than the earthquake itself can spread through the crust of the Earth~\cite{konkel2013tweets}! Similarly exciting work in forecasting with SM also exists; EMBERS is a currently deployed system for monitoring civil unrest and forecasting events such as riots and protests~\cite{ramakrishnan2014beating}. Using a combination of SM and publicly-available, non-SM data, they are able to predict not just when and where a protest will take place, but also why a protest may occur.
These findings have enticed researchers into exploring the possibilities opened by SM data, but there remain many unanswered questions. If SM is useful for detecting real-time events, can it be used to make predictions about the future? What limitations does forecasting with SM data face? What methods lead researchers to positive results with SM data? 

For all of its exciting advantages---SM platforms are global, multilingual, and cross-cultural---a deep pessimism surrounds SM data analysis \cite{ruths2014social,weller2015accepting}. SM is noisy and the data derived from SM are of mixed quality: for every relevant post there may be millions that should be ignored. Learning with SM data sometimes requires robust statistical models capable of handling massive quantities of SM data, but here too there are additional open questions about the effectiveness of such data-driven models. Consider the case of Facebook, who in 2014 launched a \emph{trending topics} feature later revealed to be hand-curated by Facebook employees~\cite{ap2016}. Facebook used an algorithm to scour the site, utilizing their own SM platform's data to detect trending topics that were then looked over by humans for quality assurance. Facebook later removed human curators from the process---following the idealized trend of SM data analysis---and relied entirely on their data-driven algorithms. Within days Facebook's system had posted libelous articles and explicit material~\cite{thielman2016}. If a SM platform as large as Facebook is unable to use its own data to detect aberrant trending topics, what are the prospects for other organizations?

Yet, in spite of anecdotes like this, researchers continue to investigate how best to make use of SM data. Preliminary results do largely show positive findings as we discuss in much greater detail below.
If SM users are reacting to and talking about events in real time, one might imagine that users are also talking about and reacting to events that they anticipate will happen in the future. This raises the interesting possibility that SM data might be useful for \emph{forecasting} events: making predictions about events that have yet to occur. Not only have researchers begun to investigate this line of questioning, earlier review articles on SM forecasting showcase early positive examples of predictive success~\cite{yu2012survey, gayo2013power, kalampokis2013understanding, o2015twitter}. Across the board preliminary studies show that SM \emph{could} be used to predict the future. At the same time, early findings have been controversial and warrant some amount of skepticism and caution~\cite{yu2012survey,gayo2013power,kalampokis2013understanding}. The field is in its infancy, methodologies are scattered, common best practices are nonexistent, and true replication of studies is near-impossible due to data sharing concerns~\cite{weller2015accepting}.

Previous reviews laid out a number of possible issues with SM forecasting and identified areas where forecasting had or had not been successful, but had little to say about what best practices researchers might follow in order to better make use of SM data. 
Identifying all of the pitfalls associated with SM data is far beyond the scope of this literature review, therefore we choose to focus on the following general questions:

\begin{itemize}
	\item [Q1:] Can SM be used to make accurate predictions about current and future events?
	\item [Q2:] Across domains, what distinguishes SM prediction successes from prediction failures?
\end{itemize}

While previous reviews were cautiously optimistic in addressing Q1, by covering a much larger body of literature, we aim to find a more comprehensive answer. We further address Q2 in order to give researchers an idea of how they might best approach their own SM forecasting tasks.
The contents of the rest of this review are organized as follows: the background section provides a general overview of SM and the interest it has generated, clarifies the meaning of \emph{prediction} and \emph{forecasting}, and describes some general challenges faced by researchers. We describe which general topics are covered in the literature review and methods and requirements for study inclusion. Next, we present our findings split by prediction topic, focusing on elections, economics, public health, threat detection, and user characteristics, addressing research questions (Q1) and (Q2) above. Further, each results section includes a table of reviewed articles that lists the primary author, topic, data source, collection method, size, primary data features, algorithmic task, success rate, and validation method of the section's constituent reviewed articles.

Our principle research questions (Q1) and (Q2) relate to how well SM data can be used to predict future (or otherwise unknown) real-world states, i.e., \emph{forecasting}. We also note that many papers focus on identifying the current state of the world, i.e., \emph{nowcasting}. Both types of papers are included in our analysis for two principle reasons. First, the state of the world is often persistent over time, meaning that current predictions may overlap with future predictions, e.g., the case of predicting a user's ethnicity. Second, predicting the future is likely to be more difficult than predicting current states because of increased temporal distance~\cite{baillie1992prediction}. The ability (or inability) of existing research to nowcast current or immediate future states is therefore an upper bound on how well forecasting states further in the future might perform.

\section*{Background}

\begin{figure}

    \centering
    \includegraphics[width=.75\textwidth]{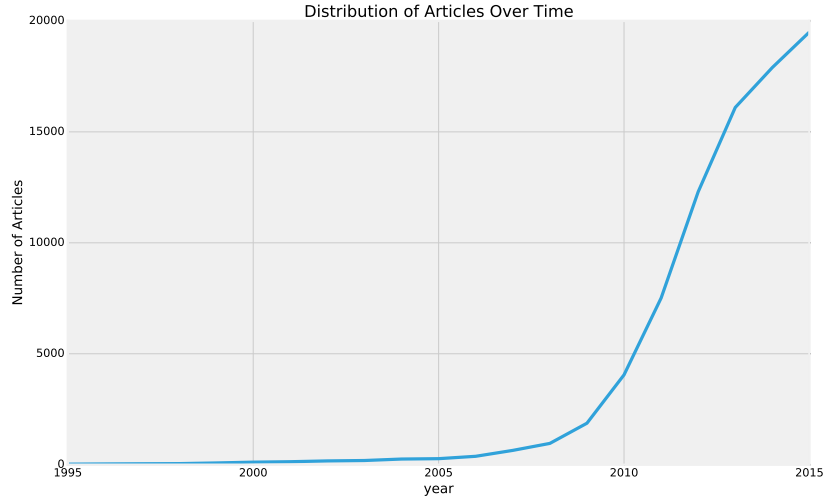}
    \caption{Number of articles published per year containing the phrase ``social media'' and the keywords ``data'' and ``prediction'' according to Google Scholar, excluding patents and case law.}
    \label{fig:refs}
\end{figure}

The use of SM data for modeling real-world events and behavior has seen increased interest since its early appearances in academic work around 2008. Fig~\ref{fig:refs} illustrates this growth, with nearly 20,000 articles having been published in 2015; meanwhile, 2016 is set to well exceed that number. 
This rise in popularity is commensurate with the newly coalescing field of computational social science~\cite{lazer2009life}. Many sociological  hypotheses were previously untestable due to difficulties in obtaining data. With the advent of SM, this is no longer the case, as myriad facets of human interaction are recorded by millions of people across the web. At the same time, this data is not always a complete cross section of what a researcher might hope to see. SM usage varies by age, culture, socioeconomic status, gender, and race~\cite{perrin2015social}. Still, positive findings and interest in the fundamental dynamics of SM platforms is a likely culprit for this exponential growth in popularity, particularly for social scientists~\cite{batrinca2015social,weller2015accepting,gayo2011limits,o2015twitter,zeng2010social,kwak2010twitter,grimmer2015we}.

\subsection*{Forecasting and Predictive Modeling}

Standard examples of physical laws and theories (e.g., Newton's Laws or the Ideal Gas Law) have provided the sciences with a means of \emph{forecasting} or \emph{predicting} natural phenomena. Specifically, given a sequence of observations related to the state of some system, \emph{prediction} entails the accurate and reproducible state estimation of that system for some amount of time into the future up to and including the present. For a simple physical system, we might use Newton's Laws to derive a model of the position and velocity for a mass on a spring (i.e., Hooke's Law). Models which build off a theoretical understanding of the underlying system are considered \emph{theory-driven} models. In many cases, however, we lack a full or even partial theoretical understanding of the underlying system. For instance, it would be quite difficult to create an entirely theory-driven model to forecast when a user is going to make their next SM post about an unforeseeable topic. \emph{Data-driven} models learn predictive relationships from data directly, for instance by looking at previous posting patterns for a user. We distinguish between theory- and data-driven models, although in practice models often incorporate aspects of both methods. Data-driven models are often used to gain insight into the fundamental laws governing the underlying system: the authors of \cite{tu2014dynamic} demonstrate how to recover or learn Hooke's Law directly from sensor data without knowledge of Newton's Laws.

\subsubsection*{Underlying Complexity of SM-based Models}

\begin{figure}
    \centering
    \includegraphics[width=.6\textwidth]{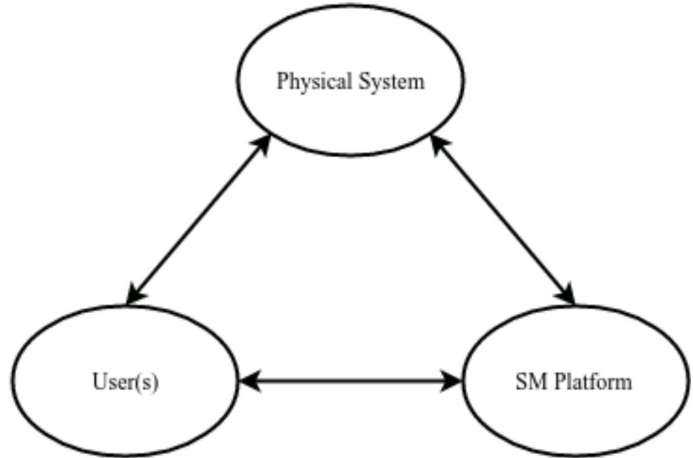}
    \caption{Interactivity of factors leading to SM forecasting ability. For example, users may observe the weather around them and post those observations on Twitter. A less common example would be weather directly effecting the social media platform, via weather-related outages.}
    \label{fig:inter_intro}
\end{figure}

It is clear that forecasting should be possible to varying degrees when there are direct causal links, as in the case of the physical systems described above, whether these links are identified through theory-driven hypothesis testing, naive data analysis, or both. If current weather patterns impact future weather, that relationship should allow for forecasting. If current behaviors impact future chance of illness, that relationship should likewise allow for forecasting. Yet in almost all cases, SM posting does not directly impact the real world system we care about and the real world system does not directly impact SM posts or behavior (which is generally the relationship being modeled for SM forecasting). Instead, physical systems in the real world and SM users interact with one another and then users interact with SM. We demonstrate these relations in Fig~\ref{fig:inter_intro}. Each arrow represents a (not necessarily causal) relationship between two systems which can be modeled, where direction matters. Forecasting can be accomplished by using theoretical knowledge to understand the underlying mechanisms which produce a link between SM behaviors and real-world outcomes, or this relationship can be modeled directly from the data. 

Prediction becomes somewhat more difficult as the gap between any two of these factors increases and their relationships becomes less \emph{direct} between one another. While many users may be influenced by or be an influencer of a stock market, for example, predicting the behavior of a stock market is already known to be an all but insurmountable task both empirically \cite{lo2002non, ball2013counting} and theoretically \cite{malkiel2003efficient}. While focused SM data analysis may yield insight into stock market behavior, SM users (corporate or otherwise) are unrepresentative of the players within a stock market and trades are often purposefully obfuscated \cite{chakravarty2001stealth}. This is to say that SM does not significantly overlap or impact a majority of the variables governing the physical system, namely a stock market in this case. 
The difficulty of establishing SM prediction for real-world events is a reflection of these underlying processes  which vary between tasks and are often only poorly understood or not taken into account.

A significant manifestation of an event on SM, however, does not appear to be a sufficient condition for successful prediction. Take for instance the 2014 World Cup; the tournament saw global SM presence representing participating teams from around the world~\cite{dowling2015social,yu2015world,radosavljevic2014large,godin2014beating,corney2014spot}. An attempt to predict match outcomes utilizing Twitter data, ~\cite{radosavljevic2014large} failed to perform better than random chance for early tournament matches, and under-performed popular sports analysis agencies' predictions beyond quarter-final matches. Much of any given team's SM traffic reflected the development of a game and general attempts to rally fan pride \cite{yu2015world}, but the SM platform's activity itself had little demonstrable bearing on the outcome of the game. Indeed, apart from a handful of sports journalists broadcasting informed \emph{a priori} analysis, the majority of fans are not directly involved in the game and are merely spectators possibly explaining poor predictive performance~\cite{radosavljevic2014large}. On the other hand, as spectators, SM users do post information which can be used to identify what is happening as a match progresses, i.e. nowcasting~\cite{corney2014spot,meladianos2015degeneracy}.

Ultimately, additional variables in complex prediction tasks increase the gap between each of the factors in Fig~\ref{fig:inter_intro}. Simpler queries with direct relevance to how users interact with SM and the physical system 
might be expected to enjoy better predictive success. Consider the case of predicting when soccer matches like the above will occur. The authors of \cite{hurriyetoglu2014estimating} achieve an accuracy of $\pm$ 8 hours up to 6 days in advance of a game. This could be attributed to the fact that attending a game directly impacts all users involved---players, journalists, and fans alike---where fans will broadcast their planned attendance and support on SM in addition to teams, players, and journalists publicizing the event \cite{corney2014spot}. 
Additionally, in some cases SM users have \emph{direct} knowledge related to the forecasting task, e.g., they do know when a game will take place well ahead of time. Such instances should be much easier to forecast than cases where any knowledge on the part of SM users is \emph{indirect}, as in forecasting match winners, where it could be argued that SM users are privy to some relevant information, but have no direct knowledge of the outcome.

\subsubsection*{The SM User-Sensor}

\begin{figure}
    \centering
    \captionsetup{justification=centering}
    \begin{subfigure}[b]{0.45\textwidth}
        \includegraphics[width=0.8\textwidth,angle=270]{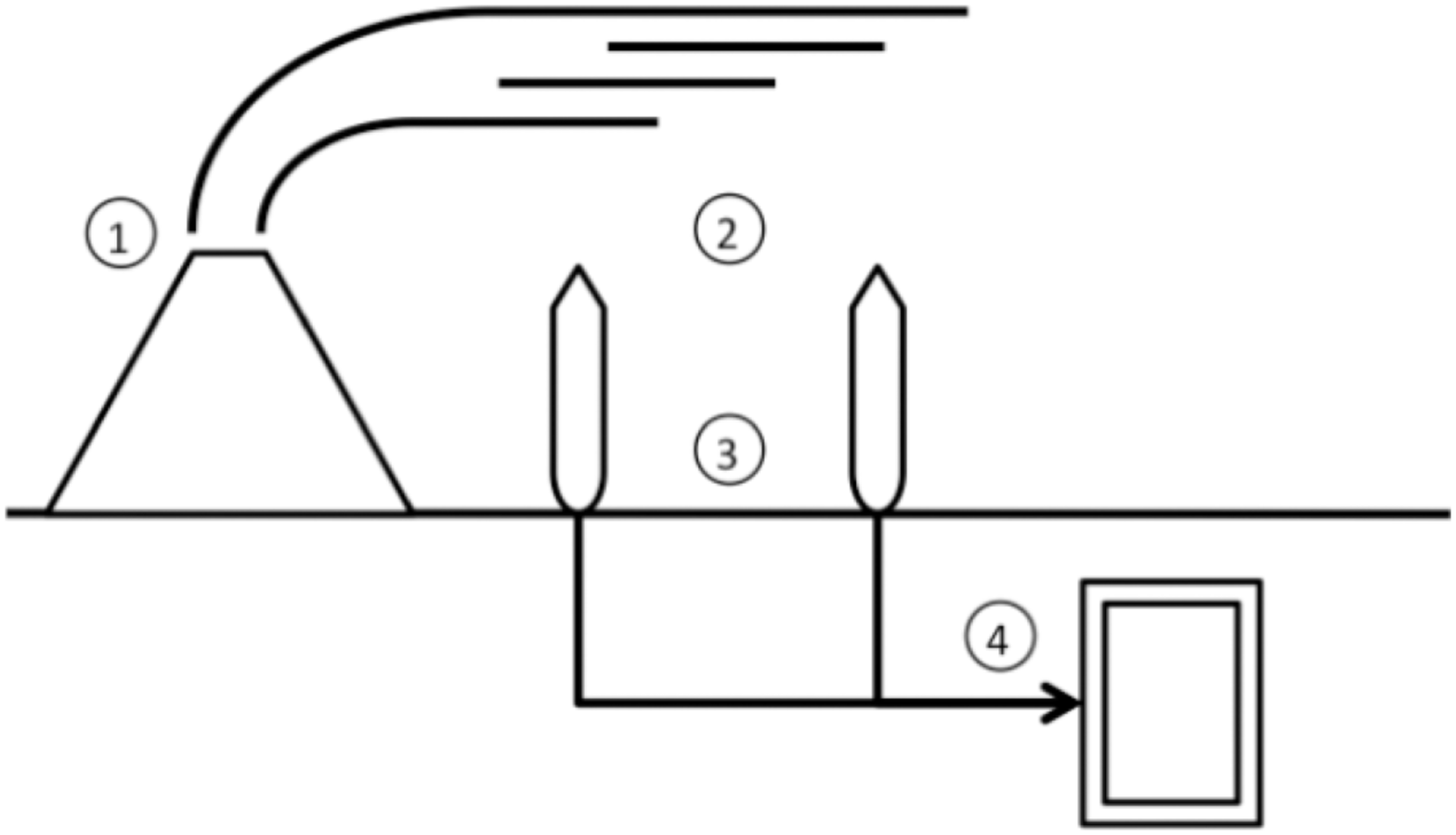}
        \caption{A traditional sensor network}
        \label{fig:sensor}
    \end{subfigure}
    \qquad 
    \begin{subfigure}[b]{0.45\textwidth}
        \includegraphics[width=0.8\textwidth,angle=270]{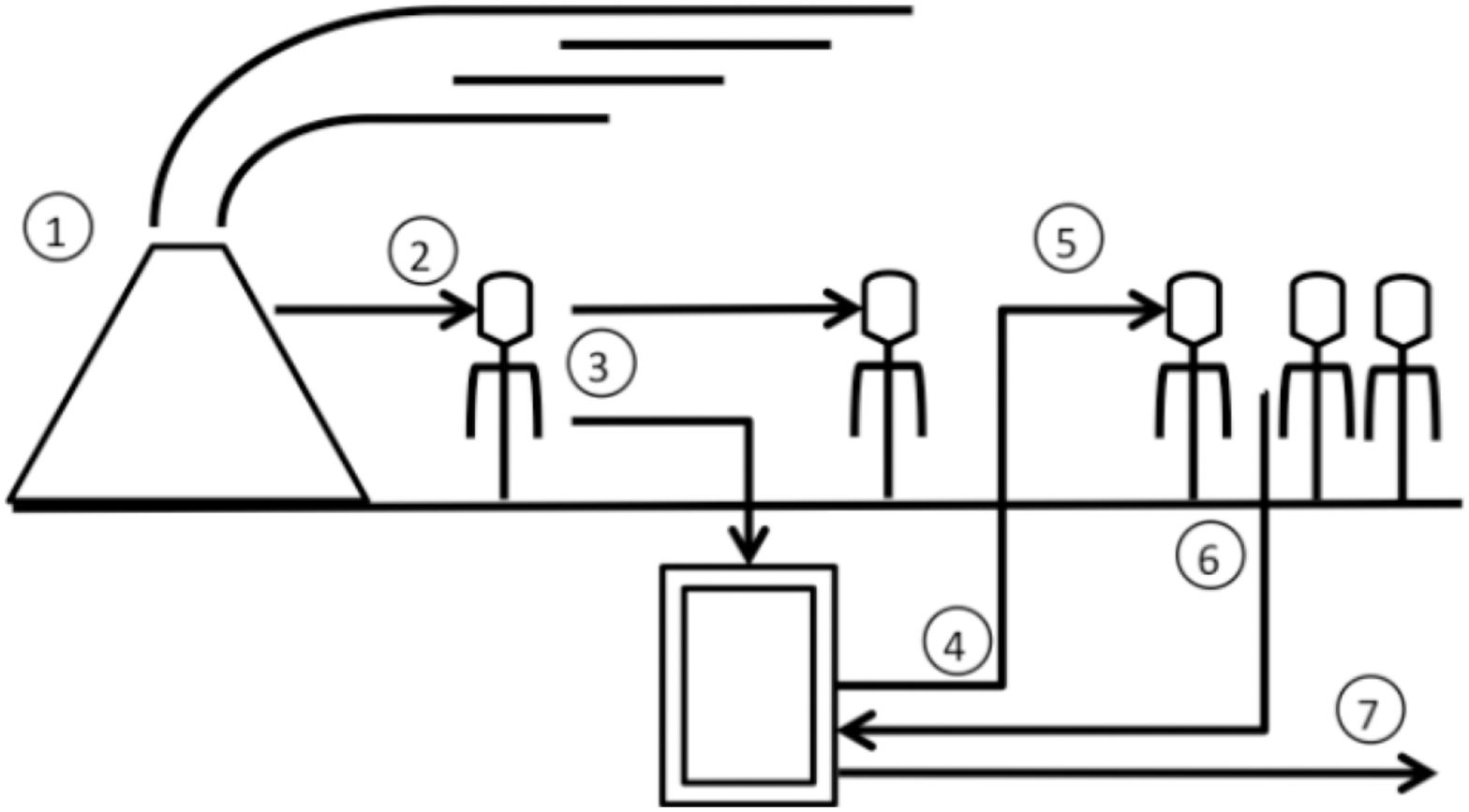}
        \caption{A SM sensor network}
        \label{fig:people}
    \end{subfigure}
    \caption{Comparison of sensor data routing: traditional vs SM sensor network, adapted from \cite{corley2013social}.}\label{fig:sensor_comp}
\end{figure}

Users themselves further complicate matters. Consider the process for detecting an event with a traditional sensor network in Fig~\ref{fig:sensor} taken from \cite{corley2013social}: 1) some physical event occurs, 2) sensors acquire a measurement, 3) the sensors record the measurement, and 4) the system stores the measurement. Although sensor readings may be correlated, the sensors do not typically interact with one another directly. 
SM users can be thought of as sensors, but the purpose of the SM platform is specifically to \emph{allow} interaction between different users. 
Consider the parallel process of event responses in a SM sensor network in Fig~\ref{fig:people}: 1) some physical event occurs, 2) user receives stimulus, 3) user communicates response, 4) system routes message, 5) other users receive message, 6) users communicate response, and 7) system routes message. 

Consider an idealized case of a traditional sensor network where one sensor is reporting false information. In such a case, the incorrect sensor's data can be compared against the data received from other sensors and because the sensors do not interact with one another, a single incorrect sensor will not cause a cascade of false information. 
On SM, however, such information cascades can and do occur. Consider the case of the 2013 Boston Marathon bombing. Immediately following the event, users on various SM platforms, in particular Reddit, began an attempt to identify the bombers. As SM users shared information with one another they mistakenly settled on Sunil Tripathi as the primary suspect. Tripathi had been missing for a month by the time the bombing took place and had in fact taken his own life~\cite{redditBostonBombing}. Because SM users react not just to outside events but also to posts from other SM users, it is possible that a user's perception of outside events is influenced by other users, essentially introducing the possibility of sensors biasing other sensors. Besides false accusations, this leaves SM sensors susceptible to other well-studied phenomena such as group polarization~\cite{yardi2010dynamic,conover2011political}.

\subsection*{Open Challenges for SM Forecasting}

This is all to say that while SM data holds tremendous potential value, its useful application is not necessarily a trivial matter. Forecasting techniques in the natural sciences, both theory- and data-driven, are relevant, but SM challenges researchers to find new ways to apply them. 
Aggregation techniques from traditional sensor networks are relevant, but SM challenges researchers to find new ways to augment them. 
Because of these difficulties, Q1 and Q2 are intricately linked. To be able to generate valid, reliable predictions (Q1) researchers must first identify the methods through which myriad challenges in SM research may be addressed (Q2). 
These difficulties include noisy data, possible biases, a rapidly-shifting SM landscape which impedes generalizability, and the need for domain-specific theory to wrap everything together. 
In order to address whether these challenges can be overcome, it is necessary to examine the literature in a systematic fashion.

\section*{Methods}

In this section we detail the methods used in our systematic literature review. We define the scope of the review, describe how studies were collected and reviewed for inclusion, and discuss potential sources of study bias. 

\subsection*{Task Overview}
While many researchers have acknowledged the potential usefulness of collecting and analyzing SM data as a way to study social phenomena, much past work has concentrated on predicting various \emph{online} aspects of social networks, in particular virality~\cite{weng2013virality} and information cascades (message propagation)~\cite{tan2014effect}\remove{, amongst others}. We restrict ourselves to reviewing work that focuses on using \emph{online} data to predict \emph{offline}---viz. physical world---events. We refer to these as `real-world phenomena.'

Previous reviews have covered similar ground but describe results without clearly identifying what aspects of each domain or methodology led to success or failure of SM prediction~\cite{yu2012survey,gayo2013power,batrinca2015social,kalampokis2013understanding}. For instance, \cite{kalampokis2013understanding} reviews the literature in 2013, and makes some very general statements regarding what techniques lead SM papers to demonstrate successful results. Unfortunately, the authors collapse these generalizations across all research domains, making it difficult to discern what techniques might be best applied in particular disciplines.  A taxonomy of predictive models used in SM research is provided by \cite{gayo2013power} and explores more specific issues by content domain, but looks at specific case-in-point examples: influenza, product sales, stock market, and electoral predictions. In a like manner \cite{yu2012survey} covers a small set of content areas and also is unable to draw strong conclusions. All three reviews come to the basic conclusion that SM should be able to make accurate predictions about current and future real-world events (Q1), but are either somewhat pessimistic or unclear about how this might be feasibly accomplished. Because of the limited scope of previous reviews, no one in the literature has adequately addressed our second primary research question (Q2): what distinguishes success from failure in studies of SM prediction across all domains? 

As can be seen from Fig~\ref{fig:refs}, a great deal of research on SM prediction has been published since the last round of literature reviews in 2013. We take advantage of this greater body of literature in order to expand our review, drawing more specific conclusions about what leads to successful predictions. In order to understand how SM prediction functions both within and across domains, we divide this review on the basis of previously well-trodden disciplines. These disciplines represent the most active research areas where SM data is being used to predict real-world phenomena. 

We first provide a general outlook for each discipline as well as the specific types of prediction tasks for which researchers in these areas use SM data. We present a table of articles including the primary author, topic, data source, collection method, size, primary data features, algorithmic task, success rate, and validation method of the section's constituent reviewed articles. We then discuss general findings from each paper, noting in particular which specific factors appear to either reduce or increase the success found by researchers in that domain (Q2). Finally, we compare the existing literature in each field across disciplines to identify which methodologies are most promising by demarcating specific examples of successful and unsuccessful research practices.

\subsection*{Data Collection}

\begin{figure}
\centering
\includegraphics[width=0.9\linewidth]{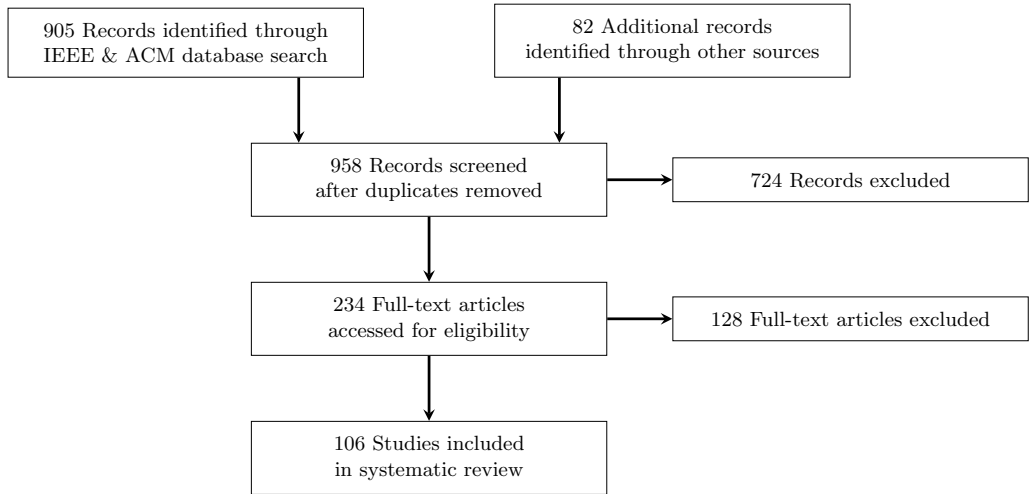}
\caption{\label{fig:PRISMA_flow}PRISMA Flow Diagram: 958 abstracts were gathered through a database search of IEEE and ACM along with a search of relevant conference proceedings. Of those, 234 full-text articles were screened for eligibility resulting in a final set of 106 studies included in the final systematic review.}
\end{figure}

To perform a systematic literature review in a research area making rapid advancements, it is important to review all topic-relevant papers regardless of their place of publication. We follow the guidelines set by PRISMA with our search outline described in Fig~\ref{fig:PRISMA_flow}~\cite{PRISMA}. We first conducted a database search in October 2016 of IEEE and ACM for articles published since 2010 which included the terms``social media" and ``prediction" as well as for ``forecasting" in the case of IEEE. This returned a total of 905 search records. We augmented this number by backward tracking references from previous literature reviews on SM forecasting as well as by searching articles from the following conference proceedings and their associated workshops from 2010 through August 2016: ACL, EMNLP, EACL, NAACL, WWW, KDD, NIPS, WSDM, ICWSM, CHI, ASONAM, AAAI, IJCAI, and SocInfo. This resulted in an additional 86 records. 

After removing duplicates this left 958 abstracts to screen. 
We included only those articles which attempted to use SM data to make real-world predictions. This included making predictions about the state of the world currently, which we refer to as ``nowcasting'', or making predictions about future states of the world, which we refer to as ``forecasting''. Articles were excluded if they did not either make predictions or attempt to discover relationships with real-world events or characteristics, e.g., speculative or theoretical articles. We purposefully excluded all articles which use SM data only to predict future SM data, e.g., research on ``virality'' which predicts the spread of SM posts on SM platforms.
After the abstract screening process, this left 234 full-text articles which needed to be assessed for inclusion. Articles were excluded for any of the above reasons as well as for failing to report concrete quantitative results, lacking real-world ground truth data, being primarily a review article, or possessing serious and obvious methodological concerns. This left a remaining 106 articles for inclusion in the systematic literature review. 

For each of the examined full-text articles, we collected information regarding the authors, topic of the study, SM platform(s) analyzed, data size (e.g. number of users, posts, images), SM features used in their analyses, the type of prediction task (e.g. classification, regression), their principle success metric and results. Because of privacy concerns, data collection methods are not always published in full and therefore where data size was not made publicly available this is noted. SM features were classified into a number of discrete categories including user metadata, n-gram counts, semantic (NLP) features, social network features, spatial or geolocation features, post volume, user behavioral features, and other non-SM features. Where multiple evaluation metrics were reported, we focus on those results primarily highlighted by the author(s) or which best represent the best level of performance achieved.

\subsection*{Study Bias}

For researchers hoping to make use of SM data for their own prediction tasks, we must qualify our observations by noting that we are unable to provide a systematic analysis of work which has not been published. Bias to publish studies with positive results necessarily taints our view of what SM can accomplish \cite{ruths2014social}. There may well be domains where SM forecasting has been attempted, failed, and  the results were not published. Because of our ignorance in these matters, we can make only reasoned assumptions about the possibility of success in domains not represented in the current review.

Further, the selection of current studies is biased in terms of which SM platforms have been studied. By far the most studied platform has been Twitter, due in large part to the ease of acquiring its data~\cite{o2015twitter}. Much of the research on Twitter may not be applicable to alternative SM platforms. Although images play a crucial role in SM they are particularly understudied and therefore we can say little about their possible predictive value \cite{wing2014hierarchical,johnson2016geography,preis2013quantifying,alanyali2016tracking,you2014eyes}.

Likewise, current research has largely focused on SM data in English and on events in the United States. It is unclear how well techniques suited toward the demographics of the U.S. can be applied to other countries, although where this has been explicitly conducted there have been largely positive results~\cite{tsugawa2015recognizing,sang2012predicting,denecke2012making}.

\section*{Results}

Research on SM forecasting spans a very wide range of topics. In order to make better sense of the existing body of literature, we split our discussion based on five general domains which have been most explored: Elections and politics, stocks and marketing, public health, threat detection, and user characteristics. For each of these domains, we discuss the existing literature in terms of its general topics and methodologies, noting particular successes and failures. We also present a detailed table which includes a number of characteristics for each study including the article's topic, data source and size, features used, the type of task (classification or regression), and their reported results.

\subsection*{Elections}

Research in election prediction has provided significant insight into the capabilities and limitations of predictive models trained using SM data. Social media platforms have allowed users to share their opinions and sentiments on a variety of topics, particularly in political discourse, and this has spurred a great deal of interest in predicting the outcomes of elections and other policy issues\cite{conover2011political}. Political forecasting is one of the first content areas to be explored with SM data, with a number of studies published by 2010\cite{williams2008political,tumasjan2010predicting,goldstein20102010,o2010tweets}, with a comprehensive meta-analysis conducted in 2013~\cite{gayo2013meta}. Given that these platforms provide a large archive of how people have talked about political and social issues, researchers have investigated the utility of this potentially useful data source in predicting and forecasting various aspects of elections and political life. In particular, research has largely focused on two specific tasks, forecasting election outcomes and now- or forecasting public opinion.

\subsubsection*{Election Outcomes}

Many polling companies spend large sums to predict the outcomes of major elections. A great deal of early work on SM forecasting focused on the predictive power of microblogs, such as Twitter, to supplement or even replace expensive polling methods. Researchers have investigated a variety of techniques ranging from extremely simplistic~\cite{williams2008political,tumasjan2010predicting,skoric2012tweets} to somewhat more complex~\cite{marchetti2012learning,ceron2014every}. Election forecasting is a very difficult task in part because major elections do not occur very frequently. Because SM is a relatively recent invention, training can only occur for a very limited number of past election cycles which may bias forecasting methods. Additionally, if a major election occurs only once per few years (e.g., 4 years in the case of U.S. presidential elections), then even legitimate predictors of a past election may well have changed in the intervening years. 

The simplest method for forecasting election outcomes is based on assuming the volume of tweets mentioning a party or candidate reflects the share of the vote that will be won~\cite{williams2008political,tumasjan2010predicting,skoric2012tweets,khatua2015can}. These models collect tweets over a period of time before an election and filter for those tweets which mention a single candidate or party running for office. It is assumed that if a candidate is mentioned in 55\% of these tweets, that they will receive 55\% of the vote and will therefore be the winner. Two of the papers using this method purport to find extremely promising results~\cite{williams2008political,tumasjan2010predicting}. In particular, \cite{tumasjan2010predicting} report that their method for predicting German election outcomes is almost comparable to traditional polls. Unfortunately, later work has cast much doubt on such simple methods. \cite{jungherr2012pirate} replicate the work of \cite{tumasjan2010predicting} showing that the model relies crucially on excluding the German Pirate Party, a new party which represented 34.8\% of mentions (almost twice that of the next most mentioned party, the CDU) but which garnered only 2.1\% of the vote. Further, they show that even when excluding the Pirate Party slight changes to the dates of data collection can lead to major changes in forecasting error. A further difficulty for volume-based approaches is mentioned by~\cite{tumasjan2010predicting}, the 3.9\% of users who tweet most heavily account for 44.3\% of all political tweets in their data. Despite the possibility of heavy bias, they make no attempt to correct for this. 
Poorer performance for the same method is reported by \cite{skoric2012tweets} and \cite{khatua2015can} who forecast Singaporean and Indian elections, respectively. While Tumasjan reports a MAE of only 1.65\%, the same method achieves 5.23\% MAE~\cite{skoric2012tweets} and 4.5\%~\cite{khatua2015can} casting further doubt on the utility of raw volume analyses.

One method for improving volume-based approaches is to take into account whether a candidate or party is mentioned in a positive or negative light. A number of studies explore whether this type of basic sentiment analysis might improve a simple volume-based approach, finding mixed results~\cite{chen2012twitter,sang2012predicting}. For instance, \cite{sang2012predicting} attempted to augment a basic volume-based approach by weighting total tweet counts based off the percentage of tweets which positively mentioned a political party. When combined with other normalization techniques (e.g., counting only one tweet per user and throwing away tweets mentioning multiple parties) sentiment improves results. Unfortunately, even after applying sentiment, the results of normalization are still worse than the basic predictions made by simple counting of mentions. The work of \cite{chen2012twitter} models U.S. Republican presidential primaries from 2012 and counts a twitter user's vote as a function of the number of positive and negative tweets mentioning a candidate. Although they do not present results for a raw count prediction, their sentiment predictions are not particularly impressive. Broken down by user demographics, most groups struggle to reach above 50\% (i.e., random) accuracy. Looking across various data collection time windows, only a single group (right-leaning Republicans) averages well above 50\%, but even then stands at 67.5\% averaged accuracy. The 2012 Republican primaries were also considered by \cite{pimenta2013comparative} who examined the difference between traditional raw volume analyses and sentiment analyses which take into account the popularity of a post. They correlate predictions from blogs, Facebook, Twitter, and Youtube against traditional Gallup polls and present two major findings. First, blogs and Facebook provided strong polling forecasts when taking into account a post's popularity. In contrast, predictions from Twitter are much poorer overall and actually decrease when taking into account retweets while Youtube predictions are the worst quality regardless. Second, they correlate their forecasts against vote totals for each candidate replicating the above findings, with Facebook and blog posts being better real-world predictors. With such a limited number of data points to evaluate against, however, it's unclear whether these positive results might be replicated elsewhere.  
Poor findings in the field of election prediction overall suggest that volume of SM posts alone, with or without sentiment analysis, is likely a poor method for predicting election outcomes.

Prediction based solely on the number of tweets mentioning a candidate is a very rough method which fails to take into account a variety of other features which might be useful in predicting election outcomes. For instance, \cite{cameron2015can} demonstrates that while the number of Facebook friends a candidate has on election day correctly predicts winners in only 16.7\% of the 2011 NZ elections, a baseline model featuring whether the candidate is an incumbent, whether the friends are of the same party as the incumbent, and similar control variables achieves accuracy of 88.1\%. Adding Facebook predictions to these control variables improves accuracy to 90.5\%. The work of \cite{ceron2014every} explores the possibility of modeling election outcomes based on tweet texts. They use a small set of hand-annotated tweets in order to estimate aggregate (rather than tweet) level sentiments. Using this method they are able to forecast the outcome of the 2012 French presidential elections roughly on par with traditional polling. They also forecast 2012 French legislative elections with a mean absolute error of 2.38 percentage points, as compared to an average of 1.23\% for traditional polls. In predicting Taiwanese elections, \cite{wang2016boosting} attempt to incorporate some notion of the popularity of SM posts. While this almost halves their prediction error, their results are still poor with an MAE of 4.0\%.  Finally, \cite{dokoohaki2015predicting} use the social graph structure of Twitter to forecast national and EU elections in Sweden. They restrict their analysis to the accounts of politicians with the idea that politicians should be more likely to win if they are influential in the SM graph, but report relatively poor correlations (EU $r=0.79$; National $r=0.65$). 

From the existing literature it is clear that elections can be forecasted using SM data, although not with the same accuracy as traditional polling~\cite{tumasjan2010predicting,ceron2014every}. Additionally, the types of features used play a large role in results. Simple volume is a very poor predictor~\cite{jungherr2012pirate} even when augmented with sentiment and taking into account the number of users rather than raw tweets~\cite{sang2012predicting}. The value of SM above and beyond simple baselines, however, may be relatively small~\cite{cameron2015can} unless more advanced techniques can be utilized~\cite{ceron2014every}.

\subsubsection*{Public Opinion}

An alternative goal for SM researchers has been to use online sentiment to nowcast public opinion, often with the goal of replicating traditional candidate approval polling. Although traditional polls are quite valuable sources of information, they are expensive and take time to gather. SM, on the other hand, can be gathered almost instantaneously, opening the possibility that SM could provide the ability to forecast ahead of polls. As with forecasting election outcomes, polling fore- and nowcasting can be built off features such as tweet volume and sentiment~\cite{o2010tweets,marchetti2012learning,saleiro2016sentiment} or word choice~\cite{ceron2014every}. 

In order to forecast both consumer confidence and presidential approval ratings, \cite{o2010tweets} gather tweets containing a small set of keywords and then measure tweet sentiment based on a previously available sentiment lexicon. Using the ratio of positive and negative sentiment on Twitter, they find a correlation both with Gallup polling on consumer confidence (released every three days) as well as with the Michigan Index of Consumer Sentiment (ICS) which is released monthly. In terms of forecasting, they explore the possibility of predicting the \emph{next} month's ICS, finding a correlation of $r=0.78$. This is worse than predicting by using previous ICS values ($r=0.80$), but incorporating both Twitter and previous ICS features improves the correlation marginally ($r=0.81$).   Correlations are also reported by \cite{o2010tweets} in comparing Twitter sentiment about President Obama with Gallup job approval ratings, but \cite{marchetti2012learning} replicates these results on a slightly different range of dates and finds much poorer performance with the same method ($r=0.22$ vs. $r=0.73$). In order to improve their results, \cite{marchetti2012learning} decide not to filter only on tweets including the word ``Obama'', instead creating a political tweet classifier. Further, rather than using a keyword-based sentiment lexicon, as in~\cite{o2010tweets}, they create a supervised classifier which learns what vocabulary is associated with positive and negative sentiment based on the emoticons used in political tweets. Correlating the resulting sentiment ratio with Gallup polls, they report a final correlation of $r=0.64$. Unfortunately, it is unclear whether this represents an advantage over simply using previous Gallup polls to forecast future poll results.

One difficulty in fore- and nowcasting public opinion using sentiment comes from the fact that not only are there a range of machine learning techniques which could be applied, but there are also any number of aggregation functions which could be used to represent sentiment. For instance, one might consider only the total number of tweets positively mentioning an entity. Alternatively, one could consider the ratio of positive to negative mentions. A wide variety of these functions are considered by \cite{saleiro2016sentiment}, who use tweets to nowcast public opinion regarding five Portuguese politicians during the Portuguese bank bailout (2011-2014). They find multiple combinations of regression algorithms and sentiment functions which all converge on a similar level of performance, MAE 0.63\%. In contrast to other work in the field, this represents a level of nowcasting performance which outperforms simply using previously published public opinion polls.

Given the conflicting results for both predicting election outcomes and polling data, does SM hold any power in predicting political outcomes? A combination of meta-analyses \cite{gayo2013meta}, literature reviews \cite{gayo2012wanted, gayo2013power}, and editorials \cite{goldstein20102010, gayo2012no, gayo2011don} have argued against the effectiveness of the predictions made above. Attempting to reproduce some of the above work, \cite{gayo2013meta, metaxas2011not} both fail to show that the proposed methods can consistently perform better than random chance. Additionally, research into election prediction has exhibited a degree of confirmation bias and is subject to the effects of heavily biased populations \cite{gayo2011don}. Indeed, in the case of Twitter, the users who choose to engage in political discourse are quite rare and focusing on these users for prediction tasks introduces selection bias in all the analyses presented.

Despite an uncertain outlook, \cite{gayo2013meta} remains hopeful that improvements can be made, and more powerful and useful models can be constructed for effective prediction in this domain. Indeed, in looking at the summary presented in Table \ref{tab:polisci}, we see that the vast majority of prior work relies on fairly simple methods, ranging from standard linear models such as linear or logistic regression, to simple keyword matching. With recent advances in machine learning models, including ensembling and neural networks, there is a great deal to explore in applying these methods to SM data for election prediction. In addition to these methodological issues, more work needs to be done on actually \emph{forecasting} these election events. That is, elections tend to be regularly scheduled, recurring events that can be planned for in terms of forming a predictive task. Launching real-time studies such as this would aid in researchers getting a more holistic picture of the state-of-the-art without overfitting a model to a validation set in a post-hoc analysis, as pointed out by \cite{gayo2012no}. Overall, this area of research has seen a great deal of investment and the limitations of past studies will likely help inform further research in this area, especially given recent interest in political polarization and its effect on SM interactions.

\begin{table}[h!]
\begin{adjustwidth}{-0.5in}{0in} 
    \caption{\bf{Summary of Studies on Political Science}}\label{tab:polisci}

\begin{tabular}{ | p{1.5cm} | p{3cm} | p{1cm} | p{2cm} | p{2.5cm} | p{2cm} | p{4cm} | }
\hline
	\bf{Article} & \bf{Topic} & \bf{Data Source} & \bf{Data Size} & \bf{Features} & \bf{Task} & \bf{Success Rate} \\ \hline \hline
	Marchetti \cite{marchetti2012learning} & Approval rating & T & 476M tweets & Semantic & Regression & $r$ = 0.71 (approval) \\ \hline
	O'Conner \cite{o2010tweets} & Approval rating & T & 1B tweets & Semantic & Regression & $r$ = 0.81 \\ \hline
	Saleiro \cite{saleiro2016sentiment} & Public Opinion & T & 239K tweets & Semantic & Regression & MAE 0.63\% \\ \hline
	Tumasjan \cite{tumasjan2010predicting} &  Election prediction & T & 104K tweets & Semantic, Volume & Regression & MAE 1.65\% \\ \hline
	Skoric \cite{skoric2012tweets} &  Election prediction & T & 110K tweets & Volume & Regression & MAE 5.23\% \\ \hline
	Sang \cite{sang2012predicting} &  Election prediction & T & 28K tweets & Volume, Non-SM &  Regression & 29\% worse than polls \\ \hline
	Cameron \cite{cameron2015can} &  Election prediction & T, F & Not specified & Volume, Non-SM & Regression & Acc. 94.6\% \\ \hline
	Dokoohaki \cite{dokoohaki2015predicting} & Election prediction & T & 130K users & Social & Regression & $r$ = 0.65 (Swedish), 0.79 (EU)\\ \hline
	Pimenta \cite{pimenta2013comparative} & Election prediction & T, F, Y, O & Not specified & Social, Volume & Regression & MAE 5.33\% (T), 1.64\% (F), 10.42\% (Y), 1.67\% (O) \\ \hline
	Wang \cite{wang2016boosting} & Election prediction & O & 27K posts & Sentiment, Volume & Regression & MAE 4.0\% \\ \hline
	Khatua \cite{khatua2015can} & Election prediction & T & 0.6M tweets & Volume & Regression & MAE 4.5\% \\ \hline
	Volkova \cite{volkova2014inferring} & Political party affiliation & T & 1K users &  N-gram, Social &  Classification & Acc.  99.9\% \\ \hline
	Ceron \cite{ceron2014every} & Election prediction, approval rating & T, O & 430K tweets & Semantic & Regression & MAE 2.4\% (election), MAE 8-10\% (approval) \\ \hline
\end{tabular}
\caption*{ T = Twitter, F = Facebook, Y = Youtube, FR = Flickr, O = Blogs, other}
\end{adjustwidth}
\end{table}

\subsection*{Stocks, Marketing, and Sales}

\subsubsection*{Stocks}

Of all SM forecasting tasks related to economics, predicting fluctuations in the stock market has been the most studied by far. Early work focused largely on predicting whether aggregate stock measures such as the Dow Jones Industrial Average (DJIA) would rise or fall on the next day, but forecasting can also involve making more detailed predictions, e.g., forecasting market returns or making predictions for individual stocks. The task is well studied outside of social media with a general consensus that forecasting is very difficult~\cite{lo2002non,ball2013counting,malkiel2003efficient}. As in the case of elections, most research has focused on using general SM sentiment to make forecasts~\cite{rao2012analyzing,oliveira2013some,oh2011investigating,bollen2011twitter,zhang2011predicting} although some papers investigate more nuanced models that learn the relationship between how individuals talk and market returns~\cite{makrehchi2013stock,chen2014exploiting}.

The simplest task for stock market prediction is deciding whether the following day will see a rise or fall in stock prices. Comparison between studies is complicated by the fact that stock market volatility, and thereby the difficulty of prediction, may vary over time periods.  High accuracy on this task was reported by \cite{bollen2011twitter}, using  sentiment analysis to achieve an accuracy of 87\%. They find that measures of ``calm'' on Twitter along with DJIA numbers from the previous three days provide the best up/down predictions. Further adding the emotion ``happy'' reduces rise/fall accuracy to 80\% but does reduce error in terms of forecasting absolute DJIA values. Importantly, they find that positive/negative sentiment analysis through the popular OpinionFinder tool~\cite{wilson2005opinionfinder} leads to no improvement over just using previous DJIA values. Their results are replicated by~\cite{martin2013predicting} and who forecast up/down movement for the French stock market with 80\% accuracy. Removing sentiment, \cite{mao2012correlating} use Tweets to forecast S\&P500 movements with much lower accuracy (68\%). Building on the techniques of~\cite{bollen2011twitter}, \cite{rao2012analyzing} uses an alternative sentiment analysis technique which is explicitly trained to learn what words on Twitter signal positive and negative sentiment. This does better than the OpinionFinder results reported by~\cite{bollen2011twitter}, achieving a rise/fall accuracy of 90.8\% on the DJIA as compared to 60\% when using historical data alone.  A similar F1 score of 0.85 was reported by \cite{oh2011investigating} when using sentiment from financial tweets rather than Twitter as a whole, while \cite{chen2014exploiting} reported slightly worse rise/fall accuracy (81\%) in forecasting the Chinese stock market. Sentiment alone is not enough for good stock market forecasting. In particular, \cite{porshnev2013machine} make use of a dictionary-based sentiment method similar to the one used by \cite{bollen2011twitter}. Although their methodology does not appear profoundly different, their up/down classification performance is much worse with an accuracy of 64\% on the DJIA, barely above the performance based on historical DJIA data reported by~\cite{bollen2011twitter}. 

Stock market forecasts can also be made in terms of predicting the actual value of a stock or index rather than simply whether it will rise or fall. Evaluating both tasks, \cite{rao2012analyzing} found that the best rise/fall accuracy does not lead to the best accuracy in forecasting stock values. In particular, while the emotion ``calm'' works well in predicting rise/fall, the addition of the emotion ``happy'' both \emph{reduces} rise/fall accuracy while \emph{increasing} more fine-grained prediction. The choice of particular emotions for any analysis is emphasized by~\cite{zhang2011predicting} who find that most emotions are poor predictors of future stock values. In their case, they find that \emph{both} positive and negative emotions tend to lead to a decrease in stock prices, perhaps linked to the effectiveness of ``calm'' for~\cite{rao2012analyzing}. Negative results are provided by~\cite{oliveira2013some} who investigate a variety of sentiment techniques to forecast stock values for nine US tech companies. No technique provides consistent improvements beyond a historical baseline, although they find Twitter is somewhat predictive of future trade volume and volatility. By also analyzing performance across a variety of tech companies, as well as composite indices, \cite{rao2012analyzing} similarly found that no method is predictive across all stocks.

More recent work has likewise found somewhat mixed results. Poor performance is reported by \cite{zhao2016correlating} who make use of a more complex, non-Gaussian statistical model. Their model forecasts the daily \% increase or decrease in the DJIA and report a 33\% root mean square error, meaning that average errors are potentially so large that either their methodology or report thereof is flawed. 
Building off of the mood analyses of~\cite{bollen2011twitter}, \cite{li2016can} introduce a variation which they use to predict actual returns rather than up/down classification. Compared to the relative success of \cite{rao2012analyzing}, the emotions used by \cite{li2016can} achieve much poorer results with the mood ``sad" providing the best correlation results but with an $r^2$ of only 0.40. 
Lastly, \cite{zimbra2015stakeholder} explore the possibility that noise in SM posts could be reduced by explicitly modeling stakeholders who affect a stock's price. To do this they draw from theoretical bases in economics and linguistics to first identify the features indicative of stakeholders (vs. other users) and then to make predictions based only off of stakeholder sentiment with $r^2 = 0.59$. 

Taken altogether, work on stock market prediction is largely mixed. While the mood-based analyses pioneered by \cite{bollen2011twitter} have largely proven valuable, slight deviations away from their methodology have seen much less success indicating that the method itself may be unreliable. Further, while there has been a great deal of success in forecasting up/down movements in the stock market, the ability to gauge how large those daily shifts will be is a much more difficult task and has correspondingly seen less success. Another concern comes from the fact that almost all work in the area has been built off of \cite{bollen2011twitter}. Whether such a method represents the best which can be achieved from SM data is quite unclear. In particular, there is little evidence to suggest that SM data is more predictive of stock markets than other readily available predictors.

\subsubsection*{Marketing}

An alternative use for SM forecasting is in the domain of marketing~\cite{jansen2009twitter}. Although there is a great deal of work predicting what kinds of topics and products might go viral~\cite{watts2007viral,leskovec2007dynamics,anderson1998customer}, we focus instead on a small sample of work which has been done in forecasting real-world outcomes. Very early work on blog posts demonstrated that blog posts about books showed little predictive power in determining whether Amazon sales would increase or decrease on the following day, but were useful in forecasting future sales spikes~\cite{gruhl2005predictive}. The authors speculate that this is because sales spikes are caused by outside events which are also captured through social media. Work by\cite{chen2015making} shows how modeling latent user properties such as personality traits like openness and neuroticism can help companies create targeted advertisements. The authors created a Twitter account which posted travel information and recommendations relevant to users who post about their travel plans. They demonstrate that aiming advertisements at users with particular personality traits improves click- and follow-rates by 66\% and 87\% respectively, representing a large increase in value for companies.

SM has also been used to study the ability of online projects to successfully crowdfund their projects through websites like Kickstarter~\cite{lu2014inferring,li2016project}. Proposed projects create a page on Kickstarter asking users to donate generally small sums to fund the project. Users are enticed with rewards based on the level of their donation such as early access to the proposed item(s). Projects have a set fundraising goal as well as a project deadline. User donations are only made if the sum of all donations is higher than the goal by the time the deadline passes. If funding reaches the goal by this deadline the project is said to have been successfully funded, while if the deadline passes without the necessary amount donated the project is said to have failed and no money is given. In theory, SM should be predictive of crowdfunding success since projects are expected to succeed based on users sharing information about the project through SM. Work by \cite{lu2014inferring} predicts whether a project will eventually succeed by making use of features relevant to the project itself (e.g., the fundraising goal), as well as social activity features (e.g., number of tweets related to the project), and social graph measures (e.g., average number of followers for project promoters). Using all of these features for only the first 5\% of the project duration ~\cite{lu2014inferring} achieved an accuracy of 76\% in predicting whether the project will be successful. Similarly, success was shown by \cite{li2016project} even when just using SM information from the first three days of the project, achieving an AUC of 0.90, reflecting very high classification performance.

\subsubsection*{Movie Ticket Sales}

Boosting movie ticket sales is an important task for marketing firms, and this has been studied specifically when marketing on social media platforms like Twitter. Indeed, the success of the 2016 film ``Deadpool'', having broken the record for the highest-grossing R-rated film of all time, is often attributed to its social media marketing strategy \cite{deadpoolarticle}.

Previous research linking SM to movie sales has demonstrated somewhat less predictive power than might have been anticipated. When \cite{mishne2006predicting} correlated box office sales for particular movies with SM information they found only moderate correlations ($r^2 = 0.29$) when using positive sentiment on blogs along with volume of blog posts. This represents a 12\% increase over using the volume of blog posts, but still is far from impressive. Better results are reported for a volume-based analyses by~\cite{abel2010analyzing}. They predict daily box office revenue for movies as well as sales rank for music albums achieving $r^2=0.74$ and $0.92$, respectively. Baseline features such as movie budget, genre, or number of theaters, which may hold greater predictive value, were not provided by \cite{mishne2006predicting}. 
Further work on movie sales was done by \cite{tang2014information} who monitored official Facebook fan pages for 50 different movies. They achieve an $r^2$ of 0.88 in forecasting total box office revenue when incorporating social network features, essentially modeling the influence of each movie's fan page, a significant improvement over using just the number of theaters showing each film ($r^2=0.68$). 

Much more positive results are reported by \cite{asur2010predicting} who correlate Twitter volume with opening weekend box office sales achieving $r^2=0.93$ with SM data alone and $r^2=0.97$ when incorporating SM with the number of theaters showing a film. Again, what is missing from their analysis is any systematic comparison of SM features with the kinds of non-SM features that would be used in any serious forecasting attempt. This limits our ability to determine the real predictive value of SM over-and-above baseline features. Further, a lack of systematic comparisons between various SM platforms makes it difficult to compare studies against one another or to know which platforms researchers should focus on in the future.

The relative scarcity of publications in this area of social media data analysis suggests that this is a rather difficult area of investigation. While many of the studies surveyed here present positive results, it is worth noting that many of these studies also opt to report correlations between model predictions and some ground-truth signal. These measures may obscure more nuanced model behavior as in more controlled machine learning experiments that use more sophisticated measures such as \emph{average precision} or ROC area-under-the-curve metrics. At a more qualitative level, the number of studies focusing on sentiment as a key indicator of stock or sales performance is striking. This presents many limitations and difficulties given that sentiment detection is still a somewhat open research area, and much past work casts sentiment as a crude distinction between positive or negative polarities~\cite{gonccalves2013comparing}.

Given the overall dearth of work in these areas, it is difficult to assess which of these areas may hold more promise over the others in terms of future research. In all cases, studies purporting to predict economic response variables by incorporating social media benefit from additional features outside social media, such as other economic indicators. This is particularly difficult when utilizing Twitter data, where the vast majority of tweets that can be collected will not mention a product of interest or the stock market. These challenges, along with a lack of deep understanding of how users interact online with respect to economic phenomena, will likely make it necessary to incorporate data outside social media in order to build accurate models in this application area.

\begin{table}[h!]
\begin{adjustwidth}{-0.5in}{0in} 
    \caption{\bf{Summary of Studies on Economics}}\label{tab:econ}

\begin{tabular}{ | p{1.5cm} | p{3cm} | p{1cm} | p{2cm} | p{2.5cm} | p{2cm} | p{4cm} | }
\hline
	\bf{Article} & \bf{Topic} & \bf{Data Source} & \bf{Data Size} & \bf{Features} & \bf{Task} & \bf{Success Rate} \\ \hline \hline
	Chen \cite{chen2015making} & Advertising & T & 5.9K users & Semantic & Regression & 66\% gain (click rate), 87\% gain (follow rate)  \\ \hline
	Li \cite{li2016project} & Crowdfunding success rate & T, F, K & 106K tweets & Metadata, N-gram, Social & Regression & AUC 0.90     \\ \hline
	Lu \cite{lu2014inferring} & Crowdfunding success rate & T, K & Not specified & Metadata & Classification & Acc. 76\%  \\ \hline
	Gruhl \cite{gruhl2005predictive} & Product sales rank & O & 300K blogs & N-gram & Classification & Acc. 63\% \\ \hline
	Bollen \cite{bollen2011twitter} & Stock market & T & 9.8M tweets & Semantic & Classification & Acc. 80\% \\ \hline
	Chen \cite{chen2014exploiting} & Stock market & SW & 256K tweets & N-gram &  Classification & Acc. 81\% \\ \hline
	Makrehchi \cite{makrehchi2013stock} & Stock market & T & 2M tweets & Semantic & Classification & 20\% gain (returns)     \\ \hline
	Oh \cite{oh2011investigating} & Stock market & O & 208K blogs & Metadata, Semantic, Non-SM & Classification & F1 0.85    \\ \hline
	Mao \cite{mao2012correlating} & Stock market & T & Not specified & Volume & Classification & Acc. 68\% (S\&P500) \\ \hline
	Porshnev \cite{porshnev2013machine} & Stock market & T & 755M tweets & Semantic & Classification & Acc. 64\% (DJIA), 62\% (S\&P500) \\ \hline
	Martin \cite{martin2013predicting} & Stock market & T & 173K tweets & Semantic & Classification & Acc. 80\% \\ \hline
	Oliveira \cite{oliveira2013some} & Stock market & T & Not specified & Semantic & Regression & $r^2$ = 0.20  \\ \hline
	Rao \cite{rao2012analyzing} & Stock market & T & 4M tweets & Semantic, Non-SM & Regression & $r^2$ = 0.95 (DJIA), $r^2$ = 0.68 (NASDAQ) \\ \hline
	Zimbra \cite{zimbra2015stakeholder} & Stock market & O & 64K posts & Semantic, Non-SM & Regression & $r^2$ = 0.59  \\ \hline
	Li \cite{li2016can} & Stock market & T & Not specified & Semantic & Regression & $r$ = 0.63 (sad), 0.49 (anger) \\ \hline
	Zhao \cite{zhao2016correlating} & Stock market & T & Not specified & Semantic, Volume & Regression & RMSE = 33.0\% (DJIA) \\ \hline
    Mishne \cite{mishne2006predicting} & Movie sales & O & Not specified & Semantic, Volume, Non-SM & Regression & $r^2$ = 0.29 \\ \hline
    Asur \cite{asur2010predicting} & Movie sales & T & 2.9M tweets & Semantic, Volume, Non-SM & Regression & $r^2$ = 0.97 \\ \hline
    Tang \cite{tang2014information} & Movie sales & F & Not specified & Social, Non-SM & Regression & $r^2$ = 0.88 \\ \hline
    Abel \cite{abel2010analyzing} & Movie \& album sales & O & 100M posts & Volume, Non-SM & Regression &$r^2$ = 0.74 (movies), 0.92 (albums) \\ \hline

\end{tabular} 
\caption*{T = Twitter, F = Facebook, SW = Sina Weibo, K = Kickstarter,  O = Blogs, other}
\end{adjustwidth}
\end{table}

\subsection*{Public Health}

Significant effort has been made in utilizing SM and other Internet data for the purpose of monitoring, predicting, and improving public health. Research on using SM for public health addresses a wide range of phenomena, including monitoring and forecasting disease outbreaks, identifying individuals in need of mental health services, and identifying specific adverse drug effects before they were discovered by the U.S. Food and Drug Administration (FDA)~\cite{chou2009social,moorhead2013new,feldman2015utilizing}.  An overview of how Internet data in general can be used in the public health domain is given by \cite{hartley2013overview}, and~\cite{bernardo2013scoping} gives a chronology of developments in utilizing SM data. Early work, as surveyed by~\cite{salathe2012digital}, identified the potential utility of incorporating SM into public health-related tasks. More recent comprehensive reviews confirm this potential while noting the lack of actual systems taking advantage of SM~\cite{grajales2014social,velasco2014social,charles2015using}.

\subsubsection*{Influenza}

Success in predicting epidemiological outbreaks was reported by \cite{corley2013social,cook2011assessing, li2016wisdom, hartley2013overview} to varying degrees.  A canonical example of sentiment and time series analysis in Twitter over the 2008-2009 influenza season in the United States was provided by \cite{corley2010using}. The authors report a high correlation between queries for curated vocabularies in Twitter data and influenza-like illness prevalence in the United States. Work by \cite{hartley2013overview} shows that an outbreak of dengue fever on Madeira Island (a Portuguese territory) was tracked in real-time using online biosurveillance techniques. Work by \cite{li2016wisdom} utilizes an agent-based model~\cite{bothos2010agent} but reports ambiguous predictive power with data collected from a purpose-built application. When used in conjunction with and validated by traditional data sources via the Center for Disease Control (data specifically from the Outpatient Influenza-like Illness Surveillance Network), Twitter data can reduce forecasting error by 17-30\% \cite{paul2014twitter} reports. Specifically, \cite{lamb2013separating} combines part-of-speech tagged, stemmed, and Amazon Mechanical Turk labeled Twitter data with external sources like Google Flu Trends to gain this increased forecasting resolution. In related work, \cite{culotta2015predicting} find that models augmented with Twitter $n$-gram and LIWC features are more accurate in predicting 20 county-level health-related statistics.

Although not expressly an SM data source, Google Flu Trends (GFT), released in 2008, has been the topic of much discussion in the literature, and is often used as a basis for comparison in Internet data-based biosurveillance models~\cite{velasco2014social}. Work by \cite{cook2011assessing} gives an early assessment of GFT's ability to predict influenza outbreaks by monitoring the search prevalence for influenza-like illness symptoms, showing promising results and supporting early excitement in epidemiological research. However, \cite{olson2013reassessing} subsequently reevaluate these results, and show that because GFT leans so heavily on correlative measures assumed to be good predictors, the models developed ultimately did not anticipate the 2009 H1N1 pandemic and severely overestimated both the 2011-2012 and 2012-2013 flu seasons. In fact, \cite{lazer2014parable} shows that GFT overestimated the number of anticipated cases in the 2011-2012 season by more than 50\%. This could be explained by a shift in public attention to influenza following the 2009 pandemic. 
As \cite{lazer2014parable} note, GFT makes the assumption that online behavior is determined by outside events (e.g. illness) but does not take into account the way online platforms shape the way users search. For instance, if a search platform suggests to a user that a query related to ``fever" or ``cough" might be flu-related, this may influence the user to continue searching for influenza-related information. These additional queries could bias GFT to believe the flu is more common or severe than it actually is.

Some doubt regarding studies which correlate SM with influenza-like illnesses is cast by the work of \cite{bodnar2013validating}. They expand upon previous work on Twitter demonstrating that high correlations may in fact be the result of questionable methodology. They replicate three studies, demonstrating that the similar, or even better, performance can be achieved using irrelevant or falsely generated data. This suggests that the mathematical models used in previous studies may be too powerful, overfitting the small amount of real-world influenza data which should lead to difficulty in generalizing the model to new data (as seen with GFT). This position is strengthened by the fact that the same models generalize very poorly when trained on one spatial region and tested on another. That is to say, a model trained on data from the US Northeast will likely perform very poorly in forecasting influenza-like illness on the West coast. 

Echoing the positions taken by elections researchers~\cite{gayo2013power, gayo2013meta, metaxas2011not}, without a better understanding of predictors in SM data, developing accurate models of external events based on SM features will continue to be quite difficult. 
There are any number of powerful models which can be used to model illness in the real world, but it is unclear how well any of these methods might generalize across space or time~\cite{bodnar2013validating}. That being said, research on disease detection and forecasting continues both for influenza-like illnesses and other diseases. For instance, \cite{zou2016infectious} apply advances in deep learning to the task of detecting infectious intestinal diseases such as norovirus and food poisoning.

\subsubsection*{Mental Health}

Infectious diseases are not the only health issue with relevance to SM. Researchers have also begun to use SM to identify or predict various mental health issues including addictive SM usage~\cite{shuai2016mining}, anorexia recovery~\cite{chancellor2016recovery}, addiction recovery~\cite{murnane2014unraveling}, distress~\cite{lehrman2012detecting}, suicidal ideation~\cite{de2016discovering}, suicide rates~\cite{won2013predicting}, and post-partum depression~\cite{de2013postpartum,de2014characterizing}, as well as depression more generally~\cite{de2013predicting,resnik2013using,tsugawa2015recognizing}.

A chief challenge in the area of mental health disorders is getting help to individuals in need. Screening for mental health problems is expensive and many disorders may make individuals less likely to seek out professional help. Social media promises a cheap, and possibly immediate, method of identifying individuals who may benefit from outreach. While current work suggests SM may hold great promise, there are also a number of limitations involved, including a lack of systematic reviews, differing methodologies, and difficulty in creating a ground-truth for model comparison.

Of all mental health disorders, depression has received the greatest attention from SM researchers. In the works of  \cite{de2013predicting,de2013postpartum,de2014characterizing,de2013social} postpartum and general depression was studied among Twitter users taking advantage of behavioral features (e.g., volume of tweets, number of replies) as well as linguistic features (e.g., positive and negative sentiment, use of pronouns). In all three studies, ground-truth for depression was measured by having each user fill out a survey on depressive symptoms. The reported studies were able to achieve accuracy rates of 72.4\%~\cite{de2013predicting}, 74.6\%~\cite{de2013social}, and 80.5\%~\cite{de2013postpartum} in identifying depressed users. This work finds that depressed users on SM can be characterized by decreased social activity as well as increased negative sentiment and use of personal pronouns. In order to validate their findings,~\cite{de2014characterizing} conducted interviews with the 165 subjects of their study. The authors found that reduced social interactivity alone explained up to 50\% of the variance in collected data. Similar results are also found for Japanese Twitter users, possibly reducing concerns that previous work on English-speaking users would not be generalizable~\cite{tsugawa2015recognizing}.

Social media data has also been used to make predictions about recovery. \cite{chancellor2016recovery} use survival analysis to examine Tumblr users who self-identify as anorexic. Even using very simplistic behavioral and linguistic features, they are able to predict recovery higher than chance. They identify specific features that predict recovery and compare these against features suggested by previous literature. 
Work by \cite{murnane2014unraveling} applies a similar technique to Twitter users attempting to overcome nicotine addiction. Their features and model are also quite simplistic but are still able to show clear, statistically-significant differences between relapsers and those who successfully quit smoking. While exploratory, their work does suggest that simple features tied to the existing domain-specific literature may contain the signal necessary for proper classification. This is supported by quantitative results presented by~\cite{de2016discovering} who attempt to predict if posters on mental health forums on Reddit will show signs of suicidal ideation. They report an accuracy of 80\% in predicting whether users will begin posting in the next few months to the subreddit \href{www.reddit.com/r/SuicideWatch}{r/SuicideWatch}, a forum for users thinking about committing suicide. Compare this to the work of~\cite{burnap2015machine} who classify posts related to suicide based on their intent. For example, some mentions of suicide indicate suicidal intent while others may be a report of suicide, condolence, or a flippant reference to the act. With seven classes, they achieve an F1 score of 0.69 which indicates relatively good classification on this task. 
A similar study focusing on nowcasting was able to use Reddit posts to identify individuals who were distressed~\cite{lehrman2012detecting}. They report an accuracy of 54.5\% versus a baseline of 30.5\% when classifying four ways based on the level of distress. 

The value of more complex models and features is demonstrated by more recent work~\cite{shuai2016mining,chancellor2016quantifying}. The work of \cite{shuai2016mining} makes use of behavioral features related to Instagram and Facebook usage in order to detect users with a social network mental disorder, mental disorders which manifest themselves in terms of ``excessive use, depression, social withdrawal, and a range of negative repercussions''~\cite{shuai2016mining}. They achieve classification accuracy as high as 92.6\% through the use of a more complex model which better takes into account changes in behavior over time. By comparing a number of machine learning techniques, \cite{shuai2016mining} are able to show the importance of choosing an appropriate model, suggesting that previous results may be particularly limited by the less advanced techniques often employed by social scientists. 
While much of the work on mental health fore- and nowcasting relies heavily on hand-curated lexical features, \cite{chancellor2016quantifying} provide an alternative by combining a statistical technique known as topic modeling with insights from clinical annotators. By identifying the topics in SM posts related to eating disorders, they are able to classify the severity of a users eating disorder with high accuracy ($F1 = 0.81$).

Taken together, existing work on mental health disorder detection and prediction suggests that SM is a valid and useful tool. Classification performance ranges from mediocre to very good with the greatest success in areas where more advanced features and models have been used. While the greatest number of papers have been published on detecting depression, existing work also relies almost exclusively on very simple behavioral and linguistic features within a logistic regression framework. Given the strong performance of ~\cite{shuai2016mining}, there may be room for improvement within this domain if researchers are willing to apply more advanced techniques over larger quantities of data.

Much of the work on diagnosing mental health disorders from SM data makes use of techniques that can be applied elsewhere. A good example of this is \cite{de2016characterizing} who makes use of topic modeling to understand users food choices based on the food items they post to Instagram. They make use of these food topics and user geolocation information in order to detect whether a particular region is a ``food desert", an area with limited access to nutritional food items. They achieve an accuracy of 80\% in this prediction task by combining both user posted information with publicly available socio-economic data from each region. 

While applying more advanced statistical techniques and machine learning is a clear area for improvement, there remain a number of methodological difficulties that future work must address. Perhaps the greatest difficulty is in comparing against ground-truth data. Diagnosing mental health symptoms can only be done by trained professionals, which makes it difficult to know whether a particular SM user has a disorder or not. As a result, researchers typically focus their work on particular users who volunteer to fill out a survey measuring these symptoms. While this technique provides researchers with a ground-truth, it biases these studies, making it difficult to know if results can be applied to all SM users.

\subsubsection*{Adverse Drug Reactions}
Information from SM has also been used to identify drug users suffering from adverse drug reactions (ADRs), negative side effects arising from pharmaceutical drugs taken as prescribed. 
Given that ADRs are typically reported on a case-by-case basis to physicians, the ability to monitor online disclosure of these reactions at a larger scale could greatly increase the ability of medical professionals to intervene and track these cases.
Most studies in this area focus on data mined from online medical forums where individuals ask questions about their symptoms~\cite{nikfarjam2011pattern,segura2014detecting,feldman2015utilizing}, although one study also evaluates ADRs using Twitter~\cite{bian2012towards}.

Detection of ADRs first requires the identification of pharmaceutical drug users. Researchers have made use of publicly available medical dictionaries in order to train classifiers, achieving reasonable accuracy. In the work of \cite{bian2012towards} a medical dictionary is used to identify drug-related tweets and they are able to identify drug users with a mean accuracy of 74\% and AUC of 0.82. \cite{segura2014detecting} achieve somewhat better results using online health forum data, achieving 87\% precision and 80\% recall.

In terms of identifying ADRs, ~\cite{bian2012towards} use a bag-of-words approach, classifying with a support vector machine (SVM) and achieve a mean accuracy of 74\% and area under the curve (AUC) of 0.74, somewhat lower than their results for drug user identification. Somewhat lower results in ADR identification were also achieved by \cite{segura2014detecting} with a precision of 85\% but much lower recall at 56\%.  A classifier was trained by \cite{nikfarjam2011pattern} using association rules based on keywords and part-of-speech tags. They also found the task difficult, reporting a precision of 70\% and recall of 66\%. Together, these studies indicate that identifying ADRs can be accomplished somewhat successfully using a wide range of features and classification techniques. After testing three different methods for extracting ADRs from both medical forum and Twitter data \cite{yates2015extracting} found best results in applying an initial filter to posts followed by a sequential model that is able to extract actual mentions of ADR-related terms from the text.

The previously mentioned studies show that SM can potentially be used to identify drug use and ADRs.  Taking this research a step further \cite{feldman2015utilizing} produced an unsupervised PMI-based classifier and used it to predict ADRs not labeled by the FDA. Their system models drug-symptom relations and classifies a symptom as an ADR if it appears more often in user comments than expected by chance. They tested their model on two case studies, cholesterol-lowering drugs and anti-depressants. Cholesterol-lowering was chosen because the class of statin drugs were relabeled by the FDA in 2011 to have cognitive impairment as a possible side effect. Among anti-depressants, Wellbutrin was relabeled in 2009 to include agitation as a possible side effect. Using their model \cite{feldman2015utilizing} were able to correctly identify both of these relations using user comments before the FDA relabelings. Among all the ADRs in their data, they achieved a high precision of 93.2\% with recall of 70.4\%.

This work shows that identification of ADRs is possible using online comments, particularly through health forums. Degree of success is mixed, likely due to varied methodology and reliance on properly integrated medical terminology databases. Still, current work suffers from a number of difficulties which might be improved. First, feature selection has generally been rudimentary, using bag-of-words~\cite{bian2012towards}, dictionary-based keywords~\cite{segura2014detecting}, and simple association rules~\cite{nikfarjam2011pattern}. The best performance, achieved by ~\cite{feldman2015utilizing}, makes use of a more complex grammatical parsing algorithm along with relational modeling of drugs, symptoms, and individuals. The success of future work likely hinges on incorporating more robust techniques from machine learning.

\begin{table}[H]
\begin{adjustwidth}{-0.5in}{0in} 
    \caption{\bf{Summary of Studies on Public Health}}\label{tab:pubhealth}

\begin{tabular}{ | p{2cm} | p{2.5cm} | p{1cm} | p{2cm} | p{2.5cm} | p{2cm} | p{4cm} | }
\hline
	\bf{Article} & \bf{Topic} & \bf{Data Source} & \bf{Data Size} & \bf{Features} & \bf{Task} & \bf{Success Rate} \\ \hline \hline
	Bian  \cite{bian2012towards} & Adverse drug reactions & T & 239 users & N-gram, Semantic, Non-SM &  Classification & Acc. 74\% \\ \hline
	Feldman  \cite{feldman2015utilizing} & Adverse drug reactions & O & 41K posts, 5.3K users & Semantic, Non-SM & Classification  &  F1 0.84 (statins) F1 0.78 (anti-depressants) \\ \hline
	Nikfarjam  \cite{nikfarjam2011pattern} & Adverse drug reactions & O & 6.8K posts & Semantic & Classification  & F1 0.68 \\ \hline
	Segura  \cite{segura2014detecting} & Adverse drug reactions & O & 400 posts  & Semantic, Non-SM & Classification  & F1 0.68  \\ \hline
	Yates  \cite{yates2015extracting} & Adverse drug reactions & T, O & 400K forum posts, 2.8B tweets & N-gram, Semantic, Non-SM & Classification & Prec. 0.59 (O) Prec. 0.48 (T) \\ \hline
	Corley et al \cite{corley2010using} & Influenza & T, O & 97.9M posts & Metadata, N-gram & Regression & $r$ = 0.63 \\ \hline
	Lamb  \cite{lamb2013separating} & Influenza & T  & 3.8B tweets & N-gram, Semantic & Regression & $r$ = 0.80 \\ \hline
	Paul  \cite{paul2014twitter} & Influenza & T  & Not specified & N-gram, Semantic & Regression & 25.3\% improvement  \\ \hline
	Bodnar \cite{bodnar2013validating} & Influenza & T & 239M tweets & N-gram & Regression & $r$ = 0.88 \\ \hline
	Zou \cite{zou2016infectious} & Intestinal disease & T & 410M tweets & N-gram & Regression & $r$ = 0.73 (Norovirus), 0.77 (Food poisoning) \\ \hline
	Zhang \cite{Zhang:2016:ESS:2896338.2897728} & Asthma & T & 5.5M tweets & N-gram & Classification & Acc. 66.3\% \\ \hline
	Chancellor  \cite{chancellor2016recovery} & Mental health & TR  & 13K users, 68.3M posts & Metadata, Semantic & Regression  & Concordance 0.658  \\ \hline
	De Choudhury  \cite{de2013postpartum} & Mental health & T & 40K tweets & Semantic, Social & Classification & Acc. 80\%  \\ \hline
	De Choudhury  \cite{de2013predicting} & Mental health & T & 2.1M tweets & Semantic, Social & Classification & Acc. 70\% \\ \hline
	De Choudhury  \cite{de2014characterizing} & Mental health & F, T & 40K tweets, 0.6M posts (F) & Metadata, Semantic, Social & Regression & $r^2$ = 0.48  \\ \hline
	De Choudhury  \cite{de2016discovering} & Mental health & R & 63K posts, 35K users & Metadata, Semantic &  Classification & Acc. 80\% \\ \hline
	Burnap \cite{burnap2015machine} & Mental health & T & 2K tweets & N-gram, Semantic & Classification & F1 0.69 \\ \hline
	Shuai \cite{shuai2016mining} & Mental health & F, I  & 63K users (F), 2K users (I) & Metadata, Social, Behavior & Classification & Acc. 78\% (I), Acc. 83\% (F) \\ \hline
	Tsugawa \cite{tsugawa2015recognizing} & Mental health & T & 209 users, 574K tweets & N-gram, Semantic, Social & Classification & Acc. 66\% \\ \hline
	Won \cite{won2013predicting} & Mental health & O & 153M posts & N-gram, Non-SM & Regression & Acc. 79\% \\ \hline
	Chancellor \cite{chancellor2016quantifying} & Mental health & I & 100K users & Semantic & Classification & F1 0.81 \\ \hline
	Lehrman \cite{lehrman2012detecting} & Mental health & R & 200 posts & N-gram, Sentiment & Classification & Acc. 54.5\%, baseline 30.5\% \\ \hline
	Culotta  \cite{culotta2014estimating} & Public health statistics & T & 4.3M tweets & Metadata, Semantic, Non-SM & Regression & $r$ = 0.63 \\ \hline
	De Choudhury \cite{de2016characterizing} & Food Deserts & I & 14M posts & Semantic, Spatial, Non-SM & Classification & Acc. 80\% \\ \hline
\end{tabular}
\caption*{ T = Twitter, F = Facebook, SW = Sina Weibo, I = Instagram, TR = Tumblr, R = Reddit, O = Blogs, other}
\end{adjustwidth}
\end{table}

\subsection*{Threat Detection}

Numerous attempts have been made to use Twitter data to detect rare or anomalous real-world events such as natural disasters, security events, and political uprisings. These types of events have garnered attention due to their implications for safety and security, and the spontaneity with which they arise. That is, unlike many of the other event types surveyed here, such as elections or the spread of influenza, these events do not occur regularly and do not have a limited set of outcomes (e.g., winning or losing an election). Rather, these events often constitute crises or disasters that an automated system should be able to detect in real-time as opposed to forecasting into the (distant) future.

Threat detection has built on more general work in event detection which aims to identify events from a stream of SM posts. For example, \cite{walther2013geo} focus on automatically identifying geographically localized events via Twitter streams using a combination of geofiltering and clustering techniques. Individual tweets are assigned to clusters based on 41 features computed from tweet text and geolocation metadata, and the resulting clusters are classified as belonging to either an ``event'' or ``non-event'' class. The authors report an \emph{F1-score} of 0.857 using a pruned decision tree. An alternative system, \emph{EvenTweet}, is proposed by~\cite{abdelhaq2013eventweet}. They incorporate geolocation features as well as similarity between keywords over a particular time range to identify tweets corresponding to real-time events. Although they tested their model specifically on soccer matches, their framework is general enough to be potentially applicable to other domains. The literature on general event detection is quite large and spans any number of specific content domains~\cite{ozdikis2012semantic,alsaedi2015identifying,alsaedi2016sensing,zhao2012identifying,diao2013unified,meladianos2015degeneracy}, therefore we focus specifically on event detection in the context of security events.

\subsubsection*{Cybersecurity Events}

Cybersecurity, otherwise known as IT or computer security, is an increasingly important area of interest for the protection of national, corporate, and organizational interests. A review of issues and state-of-the-art techniques in cybersecurity is well outside the scope of this literature review. Interested readers may find more information in recent literature reviews such as~\cite{mahmood2013security,franke2014cyber}. Cybersecurity research involving SM can be thought of in terms of two lines of work: 1) can SM be used to detect cybersecurity events on SM systems? 2) can SM be used to detect cybersecurity events affecting other systems?

Addressing the first question, ~\cite{jin2011towards} propose a method for detecting ``identity cloning accounts'' on Facebook. The goal of these attacks is to enter into actual users' friendship circles to access privileged information. Although their detection algorithm appears feasible, they are limited in evaluation due to the fact that they cannot collect information about actual attacks nor can they simulate an attack on Facebook themselves. In the work of \cite{wu2014detecting} they sought to circumvent these issues by having actual Facebook users participate in an experiment, browsing their own Facebook pages as well as browsing a stranger's. Given that intruders are likely to show different click behavior than legitimate users, they propose a detection scheme based on Smooth Support Vector Machines (SSVM). After two minutes they are able to identify intruders with an accuracy of 81.9\% and by nine minutes 92.9\%. While these results are impressive, savvy intruders might be able to circumvent the system by modifying their own behavior.

In terms of using SM to detect cybersecurity events happening outside of SM, current work has achieved less success. \cite{ritter2015weakly} explore the ability of Twitter to detect data breaches, account hijackings, and distributed denial of service (DDoS) attacks. The biggest challenge for such an attempt is the problem of data sparsity. While the volume of total tweets is quite large, tweets relevant to cybersecurity events are rare and tweets related to \emph{current} cybersecurity events are even less common. In spite of these challenges, \cite{ritter2015weakly} are able to detect these types of events to some degree. They present their results in terms of area under the curve for precision-recall plots, achieving scores of 0.716, 0.676, and 0.459 for account hijacking, data breaches, and DDoS attacks respectively.

Research on detecting cybersecurity events mirrors work on SM seen elsewhere. The most impressive results come when SM behaviors are used to detect irregular activity representative of an account hijacking as in~\cite{wu2014detecting}. Behavioral markers of the type they utilize appear to be quite powerful in prediction, but wide-scale analysis of such systems cannot be done by outside researchers. This limits, unfortunately, the usefulness of such measures. On the other hand, it is possible to detect when SM users are discussing cybersecurity events~\cite{ritter2015weakly}. Work on this topic is in its infancy, relying on topic and sentiment analysis techniques which are still under active development.

\subsubsection*{Protests, Civil Unrest, and Crime}

While cybersecurity events typically occur online, SM can also be used to detect and predict offline events. Of particular importance for threat detection are the prediction of events such as crimes, protests, and other types of civil unrest. Because mass use of SM has emerged only recently, research focuses largely on a small number of geographical regions where civil unrest has occurred in the past decade, particularly the Middle East~\cite{wei2015bayesian,boecking2014predicting,steinert2015online,kallus2014predicting} and Latin America~\cite{xu2014civil,ramakrishnan2014beating,muthiah2015planned,goode2015pricing}, although some work exists on protests within the United States~\cite{de2016social}. Whereas cybersecurity events are detected in real-time, i.e., ``nowcasting'', protests and civil unrest require coordination among individuals and research has investigated the ability not just to detect these events but also to forecast when they will occur.

Work on the Middle East has focused largely on the so-called ``Arab Spring'' in early 2011~\cite{boecking2014predicting,wei2015bayesian,steinert2015online}, although~\cite{kallus2014predicting} examines the 2013 military coup in Egypt. The work of \cite{boecking2014predicting} examines the ability of information mined from Twitter to detect and predict Arab Spring protests in Egypt within a timespan of 3 days. They use tweet content and ``follower'' relations on Twitter in order to predict events taken from GDELT~\cite{leetaru2013gdelt}, a publicly available database of political events around the world. They achieve similar results both for detection and predicting within a 3 day timespan with accuracies of 87.1\% and 87.0\%, respectively~\cite{boecking2014predicting}. They report that social relations alone achieve much better classification than tweet text content. The work of \cite{wei2015bayesian} examines a similar Twitter dataset with an alternative probabilistic graphical model. Although they do not report quantitative accuracy scores for their model, they come to a similar conclusion that including additional information beyond just tweet content increases model performance. That being said, even simple statistical relations can be used to predict protests. The work of \cite{steinert2015online} demonstrates that coordination on SM (e.g., the adoption of a small set of frequently used hashtags) is predictive of protest volumes on the following day, not just in Egypt but throughout all Arab countries majorly affected by the Arab Spring.

Little work has been done to extend the research on civil unrest detection to more developed, Western countries. This is perhaps due to the relatively decreased civil unrest within these nations. With the rise of the ``Black Lives Matter'' movement and worries about police shootings in the USA, there has been a single study attempting to extend this work~\cite{de2016social}. This research is mainly exploratory, investigating SM variables that might have been used to predict protest volumes across geographic regions. In terms of accuracy, their system captures 81\% of protest volume within 20\% of the true value, which is perhaps impressive given their use of a simple Poisson regression model using limited features.

Separate research has been conducted using ground-truth data about protests, riots, and civil unrest in Latin America focusing on a timespan between 2013 and 2014 when major protests took place in many countries. Work by~\cite{xu2014civil} and~\cite{goode2015pricing} is similar to previously examined work on the Middle East in that they both attempt to retrospectively build systems capable of detecting unrest.  Working with data from Tumblr \cite{xu2014civil} demonstrated impressive precision in event detection (95.6\%) with an average lead time of 4.8 days. 
The work of \cite{goode2015pricing} looks at Twitter data attempting to model cascading social participation, the idea that individuals are more likely to join an offline protest if they are exposed to a critical mass of online support. Their work is somewhat less successful, achieving an F1 score of approximately 80\% on their Brazilian data, but only 55\% on their Venezuelan data. It should be noted, however, that both of these models are trained on data from time periods of heavy unrest, and it is unclear whether these models would have predictive power in an implemented event detection system.

A separate line of research on Latin American protests is well worth discussing in terms of clarifying how well a fully implemented system might be capable of detecting unrest events. The EMBERS system, as described by~\cite{ramakrishnan2014beating}, is an automated event detection system which has been in place making forecasts since November 2012. EMBERS makes use of a wide variety of publicly available data including Tweets, RSS news and blog feeds, meteorological data, and Google Flu Trends among many others. While most research papers focus on a single model of event detection, EMBERS makes its forecasts by combining five separate models, only two of which has been fully investigated in separate work~\cite{muthiah2015planned,korkmaz2015combining}. They evaluate their models' forecasts based on a ``quality score'' which equally weights predictions in terms of how well they match the date of the event, the location, the event type, and the population which will be involved in the event. This results in a score from 0 to 4, with higher values indicating better predictions.  An average quality score of 3.11 is reported by \cite{ramakrishnan2014beating}, with recall of 82\%, precision of 69\%, and an average lead time of 8.88 days. Putting this into context, the results of a single model achieves a classification F1-score ranging from 0.68 to 0.95 depending on country~\cite{korkmaz2015combining}. These impressive results, not just in detecting events but in detecting specific properties of those events, demonstrate just how powerful the lines of research previously discussed might be if put into practice.

While SM may be predictive of large-scale organized protests and civil unrest, one may wonder whether the same would be true for smaller-scale criminal acts which do not have the same organizational requirements. Published work on crime forecasting is largely limited by the (in)accessibility of large-scale databases of detailed criminal records. Because of this, published research has largely focused on a single city, Chicago, which has made some of its records available for research purposes~\cite{gerber2014predicting,chen2015crime}. A wide variety of crime types are examined by \cite{gerber2014predicting}. The simplest method for predicting areas where crimes are likely to take place involves making use of historical crime data. This can be augmented by SM posts, which \cite{gerber2014predicting} analyze using a topic model, linking the learned topics to the likelihood of various crimes. Of the 25 crime types examined, 19 see a forecasting improvement by incorporating SM data. Using the same data, \cite{chen2015crime} choose to combine historical crime data with a sentiment analysis of Twitter posts as well as with current weather conditions. They report classification accuracy somewhat lower than found in \cite{gerber2014predicting} and also find that the addition of weather and SM sentiment only improves classification marginally. Further, because they do not report separate results for SM data or weather data alone, it is unclear whether this marginal improvement is in fact the result of adding SM data at all. Based on these findings, it is unclear what the best approach to making use of SM data for crime forecasting might be. While there may be some value for police organizations, it is unclear how great that value might be.

\subsubsection*{Natural Disasters}

Natural disasters represent an additional type of threat which can be detected through SM. One of the first demonstrations of SM's ability to detect events in real time came in the form of earthquake detection~\cite{konkel2013tweets}. Although earthquakes cannot be \emph{forecasted} through SM, a number of articles have demonstrated that Twitter users represent a sensor network that can potentially outperform standard seismology systems~\cite{sakaki2010earthquake,earle2010earthquake,crooks2013earthquake}. In these special cases, earthquake-related tweets can functionally predict when standard seismology equipment will detect an earthquake a few moments later. Not only can Twitter beat traditional sensors, but they can also distribute information about a quake to potentially affected individuals much faster than governmental agencies~\cite{sakaki2010earthquake}. Additionally, SM data can be used to detect earthquakes where high resolution equipment may be sparse~\cite{twitterEarthquake}. 
Other types of natural disasters are explored by~\cite{middleton2014real} who particularly focus on identifying specific locations most in need of aid. Applying their method to earthquakes, floods, and tornados, they are able to identify streets and places of interest most hard hit by disasters.

A small set of studies also exist on using SM to now- and forecast weather more generally. Concerns regarding climate as a national security issue have made these additional studies increasingly relevant in the area of threat detection~\cite{climatechange}. While this is an underexplored area, there is some evidence that SM can be used to both detect and forecast weather. For instance, the authors of~\cite{lampos2012nowcasting} use tweets from 5 UK cities to nowcast amounts of rain. Even with a relatively simple model based on weather-related keywords they are able to nowcast daily rainfall with an RMSE of 2.6mm, a 40\% error reduction versus their baseline model. Levels of air quality can also be inferred through SM data~\cite{mei2014inferring,chen2015smog,li2015using}. Looking at air pollution levels in Chinese cities,~\cite{mei2014inferring} first build a spatial model which does not take advantage of SM. This allows the model to take advantage of the fact that pollution levels across cities are correlated based on their real-world properties. Using this as a baseline, they expand to include information from Sina Weibo posts finding a 13\% reduction in prediction error. The problem of air pollution detection can also be treated as a detection task where there is either a smog event in a city or not on any given day. Taking this approach, \cite{chen2015smog} combine traditional physical sensors with SM check-in and textual data in order to estimate the population mobility and traffic conditions which might lead to smog events. Their neural network-based architecture is able to classify smog events with high accuracy using only physical sensor data, but sees an improvement in performance when adding SM data as well, indicating that even when traditional sensors are in place SM data can still add additional value. 
While the previous two works focus on textual data from SM, \cite{li2015using} make use of images posted on SM in order to monitor air pollution. While their method works very well on a small set of high quality images ($r=0.89$), on a larger set of noisy images taken from SM the method is much less successful ($r=0.41$).

Threat detection is an area of great promise with at least one detection system currently operational and demonstrating great effectiveness. Positive results through retrospective analysis utilizing a variety of modeling techniques also show the possibility of real-world usage. At the same time, it should be made clear that extremely successful results in retrospective studies, as in~\cite{xu2014civil,leetaru2013gdelt}, are unlikely to translate into similar results for actual forecasting. This is because many of these retrospective analyses formulate the supervised learning problem as only having to decide between a small set of classes, in some cases just two (e.g., event or non-event). This greatly simplifies the learning problem, but real-world systems such as EMBERS~\cite{ramakrishnan2014beating} need to make predictions not just of time and event, but also of event type, population involved, and location. These added details greatly increase the difficulty of the problem and likely require incorporating information from outside of SM, such as weather or financial news~\cite{ramakrishnan2014beating}.

\begin{table}[H]
\begin{adjustwidth}{-0.5in}{0in} 
    \caption{\bf{Summary of Studies on Threat Detection}}\label{tab:threatdetect}

\begin{tabular}{ | p{2cm} | p{2.5cm} | p{1cm} | p{2cm} | p{2.5cm} | p{2cm} | p{4cm} | }
\hline
	\bf{Article} & \bf{Topic} & \bf{Data Source} & \bf{Data Size} & \bf{Features} & \bf{Task} & \bf{Success Rate} \\ \hline \hline
	Boecking  \cite{boecking2015event} & Civil unrest & T & 1.3M tweets & Metadata, N-gram, Semantic &  Classification & Acc. 87\% \\ \hline
	Boecking \cite{boecking2014predicting} & Civil unrest & T  & 1.3M tweets & Metadata, Semantic, Social & Classification & Acc. 87\% \\ \hline
	De Choudhury \cite{de2016social} & Civil unrest & T & 29M tweets & Semantic & Regression & $r^2$ = 0.42  \\ \hline
	Kallus  \cite{kallus2014predicting} & Civil unrest & T, O & Not specified & Semantic & Classification  & AUC 0.92  \\ \hline
	Ramakrishnan  \cite{ramakrishnan2014beating} & Civil unrest & T, O  & 3B messages & Semantic, Social, Spatial, Non-SM &  Classification & Rec. 0.82 \\ \hline
	Korkmaz \cite{korkmaz2015combining} & Civil unrest & T & 500M tweets & N-gram, Non-SM & Classification & F1 0.95 (Brazil), 0.88 (Mexico), 0.70 (Argentina) \\ \hline
	Gerber  \cite{gerber2014predicting} & Crime & T  & 1.5M tweets & Semantic, Spatial, Non-SM & Classification & AUC 0.72 \\ \hline
	Chen \cite{chen2015crime} & Crime & T & 1.0M tweets & Semantic, Non-SM & Classification & AUC 0.67 (w/ SM), 0.66 (w/o SM) \\ \hline
	Wu  \cite{wu2014detecting} & Cybersecurity & F & 112 users & Behavior &  Classification & Acc. 90\%  \\ \hline
	Ritter  \cite{ritter2015weakly} & Cybersecurity & T & 14.6M tweets & Semantic &  Classification & AUC 0.72 (hack), 0.46 (DDoS), 0.68 (breach)  \\ \hline
	Avvenuti  \cite{avvenuti2014ears} & Earthquakes & T & Not specified & N-gram & Classification & F1 1.00 (magnitude $\geq$4.5 )  \\ \hline
	Sakaki  \cite{sakaki2010earthquake} & Earthquakes & T & Not specified & N-gram, Semantic, Spatial &  Clustering & MAE 3 deg. (lat./long.) \\ \hline
	Middleton \cite{middleton2014real} & Natural disasters & T & Not specified & N-gram, Spatial, Non-SM & Classification & F1 0.77 (Hurricane), 0.53 (Tornado) \\ \hline
	Mei \cite{mei2014inferring} & Weather & SW & Not specified & N-gram, Spatial & Regression & 13\% gain w/ SM \\ \hline
	Chen \cite{chen2015smog} & Weather & SW & Not specified & Semantic, Spatial, Non-SM & Classification & AUC 0.976 (current smog), 0.956 (current no smog)  \\ \hline
	Li \cite{li2015using} & Weather & O & 8.7K images & Visual & Regression & $r$ = 0.41 \\ \hline
	Lampos \cite{lampos2012nowcasting} & Weather & T & 8.5M tweets & N-gram & Regression & RMSE 2.6mm \\ \hline
\end{tabular}
\caption*{ T = Twitter, F = Facebook, SW = Sina Weibo, O = Blogs, other}
\end{adjustwidth}
\end{table}

\subsection*{User Characteristics}

While many of the areas of study previously investigated are more closely linked to predicting events in the real world, many of these tasks rely on or might be improved by accurately inferring various user characteristics. For instance, political preferences vary across sub-populations with not all populations equally represented in SM data~\cite{chen2012twitter}. If researchers could leverage inferred demographic information about users, they might be able to de-bias their models or otherwise take advantage of known demographic relationships, such as the weighting scheme employed in \cite{sang2012predicting}.

Because of the importance of predicting user characteristics, this has been an active area of research on its own, addressing a number of different prediction tasks. Potential applications have driven interest in this area of research given the possibility of major breakthroughs. For instance, security analysts might use real-world demographics to understand membership in online groups of interest, and marketers could better take advantage of ``customer segmentation'' where user locations and demographics correlate highly with purchasing certain products.

Work in each of these areas has utilized text produced by users within social media platforms~\cite{chang2012phillies, chen2013interest}, as well as images~\cite{you2014eyes}, and user metadata~\cite{harrison2015assessment, schulz2013multi, volkova2016mining} to build models for location and demographic prediction. While the studies across these areas no doubt present challenges for future work, the positive results reported indicate sustained progress in incorporating a variety of features to build ore accurate and predictive models. 

\subsubsection*{Geolocation}
Prediction of user location has been an active area of research, especially given that many social networks may contain very little geolocation metadata. The work of \cite{chen2013interest}, for example, notes that fewer than 1\% of tweets in their study contain geolocation tags. Of U.S. Facebook users in 2010, only 4\% had entered their home address in a way that could be converted to latitude and longitude coordinates~\cite{chen2013interest}. Because most users are not sharing their location information, any attempt at geolocation will necessarily be biased by the characteristics of those whose data is available.

Geolocation research can be roughly divided into two separate prediction tasks. On the one hand, location can be thought of as a static property of each user corresponding to their home address. On the other, location can be thought of as constantly shifting, such that researchers might attempt to predict where an individual is likely to be located in the future. It should be noted that we focus predominantly on mean error distances rather than medians. While many studies achieve impressive results in terms of \emph{median} errors, mean distances are often quite larger because users who are not located with high precision are often very poorly located. In terms of applicability, readers should keep in mind that an average error distance of 300 miles may seem quite large, but may still reflect a model that classifies half of users accurately within only a handful of miles.

Multiple methods can be used to predict a user's home location, but principally research has investigated the way in which social connections between users can be exploited to understand location. The work of \cite{backstrom2010find} provides the first serious attempt at this by using friendship connections on Facebook in order to locate users. They note that the likelihood of a friendship decreases as distance between two users grows. Therefore, if a user's location is unknown, the \emph{known} locations of their friends can be used to infer the user's home location. Prediction accuracy increases with the number of friends. For users with 16 or more friends with known locations, 67.5\% can be located within 25 miles, as compared to only 57.2\% accuracy for an IP-based location. By removing 75\% of known addresses, \cite{backstrom2010find} presented a much more difficult evaluation problem. Although performance decreases, they are still able to achieve an accuracy of 57.4\% within 25 miles. Work by \cite{mcgee2013location} replicates \cite{backstrom2010find} on Twitter data and report an average error of 426 miles. By incorporating the locations of user contacts they reduce that error to only 364 miles, although the top 60\% of users are located within 6 miles. A similar stance is taken by \cite{compton2014geotagging} who make use of a small set of known user locations from GPS-tagged Twitter posts in order to estimate locations for users without any explicit location information. They take advantage of the social network and by jointly estimating the locations of a large number of Twitter users ($\approx$100M) report an average error of 289km, better than previous methods~\cite{backstrom2010find,mcgee2013location}. Unfortunately, their method is unable to make location estimates for users who do not have social ties to other known-location users. Social graph-based methods are not uniformly successful. Working with Sina Weibo data, \cite{xu2014graph} find they are able to locate user home locations with an average error of 789km, much larger than that reported in other work.

An alternative early paper focused on textual differences between regions by modeling the kinds of topics used on Twitter~\cite{eisenstein2010latent}. These latent, regional topics allow for only very rough geographic estimates across the US with an average error of 900km.
By applying the task of user home location to Twitter data, \cite{schulz2013multi} annotated geolocation information for 500 randomly selected users. Their approach does not rely on social information and instead aggregates the predictions from a number of variables including tweet message, location fields, time zones, and linked web pages. Altogether their method is able to predict home location within a 100 mile radius for 79\% of their users, which is quite impressive given the noise involved in Twitter data. The work of \cite{chang2012phillies} learns a set of ``local" vocabulary in an unsupervised fashion which they use to predict which of 5,913 U.S. cities a Twitter user lives in. This approach likewise ignores social information and is able to accurately classify 49.9\% of users, an impressive feat for a purely lexical approach. 
Classifying user cities based purely on the use of specific learned keywords does poorly as reported by~\cite{kinsella2011m}. In their study they classify users from 10 distinct cities across the world but only achieve an accuracy of 65.7\%.

More recently, \cite{harrison2015assessment} show that naive spatial clustering with k-means is able to locate users home locations within 50 kilometers for 86.2\% of users. Importantly, only users with a certain number of geotagged posts could be classified. For other users, social clustering was still able to achieve 71.2\% accuracy within 50 kilometers. While most home location detection papers have focused on locating users across a large area (e.g. the entire United States), \cite{zheng2014inferring} focus on Flickr users in New York City, classifying them into $100m^2$ grids. Because users post images on Flickr of many different types, they first identify images that are likely to have been taken at the user's home and then make predictions based on the geographic information in those images, classifying with 72\% accuracy. Although all research on home location prediction suffers from similar pitfalls, including limited access to ground-truth locations, much progress has been made in recent years. While many users cannot be accurately predicted, even relatively simple clustering methods work for a majority of users. Because of limited ground-truth data, home location prediction has focused largely on a small number of users. Given the utility of social information in prediction, access to larger datasets might well lead to improved performance.

Predicting home locations is constrained by the number of users with known locations for evaluation. Predicting the location of an individual SM post is somewhat simpler in that there are many more geotagged posts. In the case of Twitter, even with only a small percentage of tweets containing location information, it is still possible for researchers to obtain millions of geotagged tweets for use in training and testing a model. Viewing the geolocation problem as one of predicting locations for individual SM posts also allows researchers to explore the ability of their models to forecast where an individuals' \emph{next} SM post will be located. This can be done by taking advantage of spatio-temporal patterns of SM users individually or by comparing individual patterns with those of similar users, incorporating social dynamics.

As with home location prediction, \cite{schulz2013multi} present initial tweet location results which are quite impressive. Their system aggregates location information from tweets, location fields, and other information such as websites and time zones mentioned in a tweet, achieving an average (mean) error distance of 1408km, but a median error of only 29.7km. This distinction between mean and median errors is driven by a small number of tweets which are very poorly localized, while the median error demonstrates that the majority of tweets are localized with high accuracy. While \cite{schulz2013multi}'s positive results rely on the combination of a variety of data types mined from SM, \cite{ahmed2013hierarchical} present a nonparametric Bayesian model which relies solely on the text of tweets. The model relies on the fact that vocabulary is shaped by geographic location and is able to achieve an average error of only 91.5km on predicting geotagged tweet locations. While different datasets make comparison difficult, this result certainly compares favorably against the 1408km average error of~\cite{schulz2013multi}, showcasing the information hidden inside SM text. Unfortunately, learning with the model is not scalable to the massive quantities of available SM data.  A much more efficient algorithm which is also based solely on text is presented by \cite{wing2014hierarchical}. In contrast to \cite{schulz2013multi} who rely on metadata specific to Twitter, this allows the model to be applied to a range of SM including Twitter, Wikipedia, and Flickr. They find promising results, with 49.2\% of U.S. tweets located within 100 miles. For English Wikipedia data that number rises to 88.9\% and for Flickr images, which are located only based on user-supplied image tags, that number is 66.0\%~\cite{wing2014hierarchical}.

These studies have all focused specifically on nowcasting the location of SM posts, and largely ignore historical patterns of individual users. This renders their techniques unsuitable for forecasting, which requires an understanding of how user locations change over time. This type of SM mobility modeling has received much recent attention as well as impressive results. A move in this direction is made by \cite{chen2013interest} who examine Sina Weibo, a Chinese equivalent to Twitter. They use tweet texts to understand individual user interests and then use these interests to predict locations. Although their model could be used to forecast, they evaluate it only on nowcasting tweet locations as in the previous studies. Direct comparison cannot be made with other studies given that they restrict their data to a single city, Beijing. They classify 72\% of tweets within 1km, but their method excludes any tweets for which no classification was made, giving their model a recall of only 15.8\%. A more useful model is presented by \cite{wang2015regularity}, which incorporates knowledge about patterns of human movement, namely the fact that individuals rarely change their patterns and that individuals with similar social backgrounds tend to share patterns. They use bus and taxi GPS data in order to model general spatial patterns of individuals in Beijing and incorporate geotagged SM tweets to understand a user's movement patterns. Their model produces a ranked list of possible locations and they present their results as Acc@topP, representing the accuracy of predictions so long as the actual location is within the top $P$ predictions. Roughly 50\% of tweets are predicting within the top 60 locations, out of a total possible of 118,534.

Recent work has expanded on the possibility of incorporating knowledge of social dynamics to predict user movements. 
The work of \cite{zhang2016gmove} specifically treats each user as a member of a latent group, where each group shares a set of movement patterns. A soft assignment for each user is made, allowing a user to belong to multiple groups simultaneously. Again, they model users on a city level both in NYC and Los Angeles. Modeling multiple groups improves accuracy by 12.7\% over a model which treats all users as members of a single group.

There are a number of hurdles to overcome in geolocation prediction. Perhaps the largest is the small percentage of geotagged data for use as a ground-truth. This is especially problematic for models that predict individual user movements, since users that do post geotagged information may well differ systematically from other SM users. A further difficulty in comparison is the variety of evaluation metrics presented, as well as differences in scale. Some studies make predictions at a country- or world-level, while others predict movements within an individual city. Errors acceptable at one level may be unacceptable at another and differences between corpora may make prediction arbitrarily easier or more difficult based on properties that are not well understood. Current efforts clearly indicate that geolocation prediction is possible. Incorporating social demographics in order to pattern users together appears to greatly improve performance~\cite{chen2013interest,wang2015regularity,zhang2016gmove}. When available, metadata are powerful sources of geolocation information~\cite{schulz2013multi}, but text alone can achieve impressive results due to differences in vocabulary across geographic regions~\cite{ahmed2013hierarchical}. If future work can leverage these initial findings and work with larger sources of data then accuracy for geolocation will likely improve greatly. 

\subsubsection*{Demographics}

While geographic location is one important hidden property of SM users, there are any number of demographic or psychological properties which can be inferred either from a user's SM relationships or posts. The two most studied hidden properties are age~\cite{perozzi2015exact,sap2014developing,ardehaly2015inferring,culotta2015predicting,volkova2016inferring,volkova2016mining} and gender~\cite{burger2011discriminating,filippova2012user,you2014eyes,sap2014developing,bergsma2013broadly,culotta2015predicting,volkova2016inferring,volkova2016mining}. Techniques for discovering these properties have improved greatly over time and reveal the power of text analysis as well as social relationships in revealing hidden properties. Further work has explored a number of other latent properties such as race~\cite{culotta2015predicting,bergsma2013broadly,ardehaly2015inferring}, education level~\cite{volkova2016mining}, political affiliation~\cite{ardehaly2015inferring,volkova2015online}, occupation~\cite{hu2016language}, income level~\cite{preoctiuc2015studying,volkova2016mining}, and even willingness to volunteer~\cite{song2016volunteerism}. While none of these tasks can be thought of as forecasting, the ability to infer the real-world properties of SM users is an important cornerstone towards expanding forecasting models. 

A wide variety of studies have investigated the ability of SM data from various platforms to infer user gender, all with positive results. Gender detection is typically formulated as a classification problem with two outcomes, ``male'' and ``female''. Initial work by \cite{burger2011discriminating} was somewhat pessimistic. Trained on text from tweets they achieved an accuracy of only 74.5\%, better than baseline (54.3\%) or human (68.7\%) performance. Incorporating a user's full name, along with other metadata, increases performance to 91.8\%, while \cite{bergsma2013broadly} report 90.2\% accuracy on their Twitter dataset. The work of \cite{filippova2012user} examines Youtube comments as a source of gender information and achieve high accuracy (89\%), although gender imbalances in their dataset lead to a much higher baseline performance (70\%). Importantly, they note that inference is much easier for individuals who tend to associate within their own gender (94\%) than for users who associate with the opposite gender (47\%). This implies that gender cues in language are used differently based on social factors and may inherently limit the accuracy achieved from text alone. In spite of this issue, text-based techniques have made large strides. The work of \cite{sap2014developing} presents a weighted lexicon-based approach that achieves 91.9\% accuracy on Facebook and 90.0\% accuracy on Twitter data. The work of \cite{volkova2016inferring} reports their results in terms of receiver operating characteristic (ROC) analysis, scoring 0.90 for text alone and 0.95 when sentiment analysis is also incorporated. Even without using text directly, gender can be inferred to some degree. The work of \cite{you2014eyes} uses image-based topics from Pinterest to achieve 71.9\% accuracy, while \cite{volkova2016mining} achieve an AUC of 0.76 using inferred user interests. The work of \cite{culotta2015predicting} infers gender by fitting users to match the demographics of websites they follow on Twitter and similarly achieve an F1-score of 0.75, indicating that gender can be inferred even when text is not available. 

Age detection is also a useful tool for SM prediction tasks. While research is consistent in attempting to infer only two gender categories, work on age detection is split between inferring age in terms of integer values (e.g., years) or as a regression task versus breaking age into brackets for multi-category classification. For instance, \cite{peersman2011predicting,volkova2016mining,volkova2016inferring} attempt to infer whether a user is above or below 25 years in age. The high F1-score of 0.88 reported by \cite{peersman2011predicting} is limited by the fact that they train a model using only word n-grams based on Dutch SM posts. While accurate, such methods are known to be brittle and do not generalize well to out-of-domain data such as posts from other SM sites. An AUC of only 0.66 is reported by \cite{volkova2016mining}, much lower than the score of 0.90 they achieve for gender detection. Age detection can be improved by adding more detailed sentiment analysis as in~\cite{volkova2016inferring}, who improve their AUC score to 0.83, still much lower than the score of 0.95 for gender. This difficulty is reflected as well in~\cite{ardehaly2015inferring} whose model based on general demographics performs very poorly on the above/below 25 classification task. They find best performance for a simple logistic regression which achieves an accuracy of 83.3\%. The work of \cite{culotta2015predicting} also attempts to use outside demographic information to improve performance and find somewhat more promising results. They split age into brackets of 10 years (e.g., 25-34, 35-44) and report correlation coefficients between inferred and actual ages in each bracket. They find best performance for ages 18-24 ($r=0.78$) and worst for ages 35-44 ($r=0.55$). Both \cite{sap2014developing} and \cite{perozzi2015exact} report mean errors in terms of years. The work of \cite{sap2014developing} reports a correlation coefficient of 0.83, better than \cite{culotta2015predicting}, with a mean error of 3.42 years for Facebook users and 3.76 years for bloggers.  Network analysis between friends on the Slovenian SM platform POKEC is used by \cite{perozzi2015exact} to infer age. Their method has lower accuracy ($r=0.70$, MAE = 4.15 yrs) 
but their results rely on only 5\% of social network users having a known age. Instagram tags and profile features are investigated by \cite{han2016teens} who report good accuracy 79.5\% when making use of both sets of features. Additionally, they evaluate their models on a completely held out set of users, ensuring the generalizability of their method.

Beyond gender and age there are a number of other demographic variables which might be useful features for forecasting. The work of \cite{volkova2016mining} refers to these properties as \emph{psycho-demographics}. While there are any number of attributes which could be analyzed, we focus our review here on six properties: ethnicity, income, education level, occupation, political party affiliation, and willingness to volunteer. 
These six attributes give an idea of what can be accurately inferred from SM data both in terms of traditional demographics, but also in terms of behavioral patterns. Due to the sparsity of work on any particular topic, we make no general claim as to the feasibility of making specific psycho-demographic inferences and instead argue that psycho-demographics appear to be detectable generally with a moderate level of accuracy.

For example, consider the case of detecting ethnicity. Early work demonstrated a promising accuracy of 81.3\% across 4 ethnic categories, but relied on having a user's first and last names which cannot always be assumed from SM data~\cite{bergsma2013broadly}. Later work showed that a much lower level of accuracy ($\sim60\%$) is possible without any kind of text data or names, based solely on matching user demographics against the demographics of websites they follow on Twitter~\cite{culotta2015predicting}. The work of \cite{ardehaly2015inferring} incorporates general demographic trends including a user's first name and reports accuracy of 82.3\%. Research on inferring user income has seen similar results. By examining a number of features to infer user income, including other psycho-demographic traits, \cite{preoctiuc2015studying}  achieves a correlation of $r=0.633$ with a mean absolute error (MAE) of \pounds9535. The work of \cite{volkova2016mining} presents their results only in terms of AUC, making comparison difficult, but they achieve an AUC of 0.67 using only inferred user interests, and an AUC of 0.73 using more traditional text analysis. In the same study, \cite{volkova2016mining} also model education levels with slightly better results, AUC = 0.70 for user interests and 0.77 for text analysis. Political party preference can also be treated as a latent psycho-demographic attribute, classifying users as either likely Republican or Democrat voters. The work of \cite{volkova2015online} reports accuracy of 81\% when using user text and neighbors. 
Occupation prediction shows similar moderate inference success, with an F1-score of .78 across 8 broad occupation labels~\cite{hu2016language}. Willingness to volunteer proves to be somewhat easier to predict using data from the SM platform LinkedIn, with an F1-score of 0.87~\cite{song2016volunteerism}, and increases to 0.899 when fusing multiple SM source \cite{jia2016fusing}.

Inference for user pyscho-demographics shows a wide range of levels of success. Social media \emph{can} be used to infer user attributes but the level of success depends on the type of data used and the difficulty of the inference task. The highest level of performance is seen in predicting gender, where text analysis of SM posts has reached similar levels of achievement to models based on user first and last names~\cite{volkova2016inferring}. At the same time, text alone is limited based on the fact that some users systematically pattern with the opposite gender~\cite{filippova2012user}. Detecting other user traits shows much more moderate levels of performance, but the sheer variety of traits which have been successfully inferred demonstrates the amount of information hidden in SM behavior. Because psycho-demographics can be used to improve performance on actual prediction tasks, even moderate success in this area is promising. 

\begin{table}[H]
\vspace*{-.8\baselineskip}
\begin{adjustwidth}{-0.5in}{0in} 
    \caption{\bf{Summary of Studies on User Characteristics}}\label{tab:userchar}
    \vspace*{-.3\baselineskip}
\begin{tabular}{ | p{2.4cm} | p{2cm} | p{1.2cm} | p{2.2cm} | p{2.9cm} | p{2cm} | p{3.6cm} | }
\hline
	\bf{Article} & \bf{Topic} & \bf{Data Source} & \bf{Data Size} & \bf{Features} & \bf{Task} & \bf{Success Rate} \\ \hline \hline
	Perozzi \cite{perozzi2015exact} & Age & O & 1.6M users & Social & Regression & $r^2$ = 0.49 \\ \hline
	Zhang \cite{zhang2016your} & Age & T & 55K users & N-gram, Social & Classification & F1 0.81 \\ \hline
	Han \cite{han2016teens} & Age & I & 20K users & Semantic, Social, Behavioral & Classification & Acc. 78\% \\ \hline
	Peersman \cite{peersman2011predicting} & Age & O & 1.5M posts & N-gram & Classification & F1 0.88 \\ \hline
	Ardehaly \cite{ardehaly2015inferring} & Demographics & T & 18M tweets, 2.7M users & Metadata, N-gram, Social & Classification & Acc. 75\% (age), 82\% (race)\\ \hline
	Bergsma \cite{bergsma2013broadly} & Demographics & T & 168M users & N-gram, Social & Classification & Acc. 90\% (gender), 85\% (race) \\ \hline
	Culotta  \cite{culotta2015predicting} & Demographics & T, O & 46K users & Social, Non-SM &  Classification & F1 0.75 (gender), 0.69 (ethnicity) \\ \hline
	Volkova  \cite{volkova2016inferring} & Demographics & T & 24.9M tweets & Semantic, Social &  Classification & AUC 0.83 (age),  0.95 (gender) \\ \hline
	Volkova  \cite{volkova2016mining} & Demographics & T  & 4K users & N-gram, Social &  Classification & AUC 0.66 (age), 0.90 (gender) \\ \hline
	Burger \cite{burger2011discriminating} & Gender & T & 213M tweets, 18.5M users & Metadata, N-gram & Classification & Acc. 92\% \\ \hline
	Filippova \cite{filippova2012user} & Gender & Y & 6.9M users & N-gram, Social & Classification & Acc. 90\% \\ \hline
	Sap \cite{sap2014developing} & Gender & T, F, O & 75K users & N-gram & Classification, Regression & Acc. 92\% (gender) $r$ = 0.83 (age) \\ \hline
	You \cite{you2014eyes} & Gender & T, P & 243K images & Semantic & Classification & Acc. 72\% \\ \hline
	Li \cite{li2014discriminating} & Gender & SW & 25K users & N-gram, Semantic & Classification & Acc. 94.3\% \\ \hline
	Eisenstein \cite{eisenstein2010latent} & Geolocation & T & 9.5K users & Semantic & Regression & MAE 900km \\ \hline
	Ahmed \cite{ahmed2013hierarchical} & Geolocation & T & 573K tweets & Semantic, Spatial & Clustering & MAE 91.5km \\ \hline
	Backstrom \cite{backstrom2010find} & Geolocation & F & 2.9M users, 30.6M edges & Social, Spatial &  Regression & Acc. 57\% \\ \hline
	Chang  \cite{chang2012phillies} & Geolocation & T  & 4.1M tweets & N-gram, Spatial & Classification & MAE 509mi \\ \hline
	Chen  \cite{chen2013interest} & Geolocation & SW & 1.1M posts & Semantic, Spatial & Classification & Acc. 70\% \\ \hline
	Harrison \cite{harrison2015assessment} & Geolocation & T  & 580K tweets, 245K users & Spatial & Clustering & Acc. 86\% \\ \hline
	Schulz  \cite{schulz2013multi} & Geolocation & T & 80M tweets & Metadata, Semantic, Non-SM & Regression  & MAE 1408km \\ \hline
	McGee \cite{mcgee2013location} & Geolocation & T & 73M users & Social, Spatial & Regression & MAE 364mi \\ \hline
	Compton \cite{compton2014geotagging} & Geolocation & T & 111M users & Social, Spatial & Regression & MAE 289km \\ \hline
	Xu \cite{xu2014graph} & Geolocation & TW & 2M users & Social, Spatial & Regression & MAE 783km\\ \hline
	Kinsella \cite{kinsella2011m} & Geolocation & T & Not specified & Semantic & Classification & Acc. 65.7\% (baseline 40.3\%) \\ \hline
	Wang \cite{wang2015regularity} & Geolocation & SW, O  & 7.3M check-ins & Spatial, Non-SM &  Classification & Acc. 45\% \\ \hline
	Wing \cite{wing2014hierarchical} & Geolocation & T, W, FR & 38M tweets, 864K posts & N-gram, Spatial & Classification & Acc. 49\% (T), 89\% (W) \\ \hline
	Zheng \cite{zheng2014inferring} & Geolocation & FR & 48K images, 192 users & Spatial, Visual & Classification & Acc. 72\% \\ \hline
	Zhang \cite{zhang2016gmove} & Geolocation & T & 1.3M tweets & Semantic, Spatial & Classification  & Acc. 50\% (LA), 38\% (NYC) \\ \hline
	Preoctiu  \cite{preoctiuc2015studying} & Income & T & 10.8M tweets & Metadata, N-gram, Semantic & Regression & $r$ = 0.63 \\ \hline
	Hu \cite{hu2016language} & Occupation & T, L & 9.8K users & N-gram, Semantic & Classification & F1 0.78 \\ \hline
	Volkova \cite{volkova2015online} & Pol. party & T & 300 users & N-gram, Social & Classification & Acc. 81\% \\ \hline
	Song  \cite{song2016volunteerism} & Volunteerism & T, F, L & 5.9K users & Metadata, Semantic, Social & Classification  & F1 0.87 \\ \hline
\end{tabular}
\caption*{T = Twitter, F = Facebook, W = Wikipedia, SW = Sina Weibo, TW = Tencent Weibo, L = LinkedIn, P = Pinterest, I = Instagram, FR = Flickr, Y = Youtube, O = Blogs, other}
\end{adjustwidth}
\end{table}

\section*{Discussion}

\subsection*{Can Social Media be Used to Predict the Future?}

Having thoroughly reviewed the literature in previous sections, we find strong evidence to support the notion that SM can be used not just to detect current real-world events, but also to make accurate forecasts into the future. Great strides have been made in the literature since the last set of reviews on SM prediction~\cite{yu2012survey,gayo2013power,kalampokis2013understanding}, and in that time the cautious optimism of previous reviews has largely been borne out. Positive results have been found in every area examined, although the degree of success is heavily moderated in part by a number of factors which we will later address (Q2). While the reviewed literature likely suffers from publication bias, failing to include many studies which found no predictive power but were not published, the fact that so many positive results have been published indicates that in some cases SM does carry great predictive power.

While there is no doubt that progress has been made, SM forecasting largely faces the same set of challenges which had previously been identified. 
First, any predictive signal in SM is surrounded by large quantities of noise~\cite{baldwin2013noisy,eisenstein2013bad} and extracting meaningful signal is no trivial task. 
Further, SM data is biased~\cite{ruths2014social,perrin2015social}. Although we can think of SM users as sensors of the real world~\cite{dowling2015social,corley2013social,sakaki2010earthquake}, it is well known that SM users are not representative of the total population~\cite{perrin2015social} and although many SM posts describe what is going on in a person's daily life~\cite{java2007we}, these posts are not necessarily representative of everything going on in the world around the user. 
As noted by \cite{yu2012survey,kalampokis2013understanding,gayo2013power,weller2015accepting}, research in SM is marred by issues of generalizability, research which makes accurate predictions on one data set may not prove useful in another. 
This issue is then compounded by the use of powerful data-driven modeling techniques without the use of any domain knowledge to link predictive power to underlying mechanisms. 
This is especially important given that many of the tasks researchers hope to predict are fundamentally complex phenomena.

The good news is that researchers in many different areas have identified at least partial solutions to many of these problems. In spite of these difficulties, they have achieved moderate accuracy of forecasts across a wide range of predictive tasks. The bad news is that achieving positive results is not necessarily straightforward. Addressing these major problems may require explicitly modeling for user biases, applying complex data-driven models, training on varied data sources, and incorporating domain-specific knowledge and theory into the modeling process. In the following section, we tackle each of these difficulties in turn, identifying studies that illustrate best practices in SM prediction.

\subsection*{What leads to Social Media Prediction Success?}

Given the variety of results presented, it is necessary to identify general trends which differentiate SM prediction success from failure. Previous literature reviews have had very little to say on the topic. The work of \cite{kalampokis2013understanding} notes that studies which use advanced techniques for filtering SM data based on keywords are more likely to find predictive success than those who use simpler filters. Further, they find that sentiment analysis techniques were highly controversial, in some cases lending no predictive power and in others proving highly useful. Based on an alternate review of the literature, \cite{gayo2013power} predicted that data-driven statistical models would increasingly find a place in SM forecasting, which has been the case.

With the explosion in research on SM forecasting over the past few years, we can make somewhat more specific statements regarding what practices researchers have used in order to find positive results. We identify four major issues which SM researchers must confront and make suggestions regarding best practices for each. We discuss each issue and best practice in detail below, presenting concrete examples demonstrating their importance for future work. The identified best practices include:

\begin{enumerate}
\item Applying appropriate techniques to overcome noisy data
\item Explicitly accounting for SM data biases
\item Learning from heterogeneous data sources 
\item Incorporating domain-specific knowledge and theory
\end{enumerate}

\subsubsection*{Overcoming Noisy Data}

While one advantage of SM data is the large quantity of data generated by users, researchers should also keep in mind that not all of that data will be useful for any particular forecasting task. Consider that in their study of adverse drug reactions~\cite{bian2012towards} find a total of 239 potential cancer drug users narrowed down from a total set of 2 \emph{billion} tweets. While this may be an extreme example, the large degree of noise---specifically information unrelated to the prediction task---represents a significant obstacle which SM researchers must overcome.

 Data filtering as an important step the data analysis process was identified by\cite{kalampokis2013understanding}. Of the studies they review, papers which manually selected keywords for filtering data supported SM predictive power only 50\% of the time. In contrast, every paper that used statistical algorithms to select keywords automatically found positive SM forecasting results. The work of \cite{tufekci2014big} provides a possible explanation for this, noting that pre-selected keywords such as hashtags, which might at first glance appear reasonable, can easily lead to poor real-world predictions. For instance, forecasting protest volumes based on hashtag usage would have led to very poor predictions in the case of Turkey's 2013 Gezi Park protests. While hashtag use dropped sharply, protest volumes and talk on SM related to the protests actually increased, because the protests were so large users no longer felt the need to use hashtags to coordinate~\cite{tufekci2014big}. This highlights the danger of focusing heavily on individual hashtags or keywords within a SM landscape which changes rapidly~\cite{weller2015accepting}. Without a principled method for filtering data, researchers risk creating models that fail due to minor shifts in word usage.

A further challenge for filtering comes from studies that make now- and forecasts based on a set of users who have been filtered for specific qualities. This type of filtering can bias predictive systems based on what was filtered~\cite{tufekci2014big}. For instance, consider the work of~\cite{volkova2014inferring,volkova2015online} who predict political party preferences for users who are either self-labeled as Democrats or Republicans or who follow exclusively Democrat or Republican candidates on Twitter. Filtering in this fashion ensures that ground-truth labels are more accurate, but also creates a bias since most Twitter users neither self-label with a political party or follow political candidates~\cite{colleoni2014echo}. 
Another example of poor user filtering comes from \cite{zhang2016your} who infer a user's age from Tweets wishing them a happy birthday while mentioning their age. Because older users rarely have their exact age mentioned, their model performs well only for users aged 14 -- 22.

An alternative approach is to use statistical techniques in order to identify signal within the massive amount of noise generated by SM without overtly removing data as in the case of filtering. Models capable of dealing with such large amounts of data can take advantage of non-obvious relations. For instance, \cite{lampos2012nowcasting} use weather-related keywords in order to predict rainfall. This allows them to learn relations between users tweeting about words such as ``rain'', ``sun'', and even indirect words such as ``beach'' or ``tv'', which might indicate activities individuals are engaging in during a sunny or rainy day. Because they rely on a fixed list of keywords, the scope of these indirect signals is quite limited. While many users tweeting the word ``fireplace'' might be a useful cue to colder weather, because it is not included as a keyword it is effectively ignored.  
A much more rigorous approach is examined in~\cite{lin2016does} who nowcast user stress levels based on a corpus of one billion tweets. Using a neural network-based architecture, they include both a set of features using stress-based keywords, but also learn from every tweet based on meanings learned from all of its words. They find that in isolation the keyword-based features are more effective than learned tweet meanings, but the combination of both significantly improves nowcasting results.
In a similar fashion, \cite{volkova2016inferring} build upon their previous work in demographic inference by removing the need for user or keyword filtering. This allows them to learn relations between user demographics and Twitter data without having a set of biased users as in~\cite{volkova2014inferring,volkova2015online}.

Noise may also be reduced in SM text data through the use of various natural language processing (NLP) algorithms.
 In particular \cite{baldwin2013noisy} examine types of noise in SM data and the effect that various NLP algorithms have on reducing it. They conclude that while traditional NLP methods developed for non-SM data are not ideally suited for SM text, they should still be generally effective.
In practice, these signal extraction methods, which do not rely on keywords, are not always perfect. Consider the case of ~\cite{Zhang:2016:ESS:2896338.2897728} who use a number of NLP techniques to reduce linguistic noise and then train a classifier to identify asthma-relevant tweets. This extraction technique does not rely on keywords and significantly improves correlations between tweets and asthma prevalence as well as monthly hospital visits. On the other hand, the method actually \emph{reduces} correlation between tweets and daily hospital visits. Even with advanced statistical models, they have difficulty forecasting whether they will see a ``high'' or ``low'' level of hospital visits either in the next day or week, reporting only 66\% accuracy for the following day. As~\cite{eisenstein2013bad} describe in great detail, standard methods for extracting signal from text are often poorly suited to online SM data. Researchers focused on SM data should therefore be wary of blindly applying such methods which were originally developed for text written by traditional media or scholarly sources.

While SM data is inherently noisy, researchers have found a number of techniques to reduce this noise and extract the signal necessary for various now- and forecasting tasks. In some cases, simple keyword or hashtag filtering is sufficient, but researchers should be aware filtering is best done in a principled fashion~\cite{kalampokis2013understanding}. Keywords chosen based on researcher intuition may be fragile, both removing possibly important information and focusing on signals which may change in importance as particular words and hashtags rise or fall in popularity~\cite{tufekci2014big,weller2015accepting}. Alternatively, statistical techniques for signal detection may be utilized without filtering. These methods automatically infer which signals are important, bypassing the need for potentially biased researcher judgments, but they have to confront the issue of noise, which may make learning difficult. In many cases, statistical methods have shown great promise, but powerful statistical models do not, in and of themselves, guarantee predictive power.

\subsubsection*{Accounting for Data Bias}

As noted in our discussion of study bias, the use of SM data to predict real-world attributes introduces a host of biases with which researchers must contend~\cite{tufekci2014big,gayo2013power,perrin2015social,duggan2015social,ruths2014social}. Not only do SM users differ from the general population~\cite{perrin2015social,duggan2015social}, but the content of SM posts also may not reflect every aspect of the real world~\cite{ruths2014social}.
These issues have been widely discussed within the literature and we broadly consider them under the name of \emph{SM data bias}.

A recent Pew Research Center study \cite{perrin2015social} shows that, apart from age, general SM adoption is largely commensurate with the growth of general Internet use among U.S. adults and representative of the U.S. population by both gender and ethnicity. In a separate Pew Research Center study \cite{duggan2015mobile} however, we find that average SM usage is not as consistent across demographics on a \emph{per-platform basis}. This is especially relevant when SM research is constrained to a single platform (which we often find in past studies, as shown in the tables above). 
As of 2015, among all Internet users, women are better represented than men on Facebook, with 77\% of Internet-using women being on Facebook versus 66\% of Internet-using men. Pinterest usage is considerably more shifted toward female than male users (44\% versus 16\%). Conversely, men are more likely to be found making use of discussion platforms such as Reddit or Digg, at 20\% of male versus 11\% of female Internet users. Platforms like LinkedIn and Twitter see much closer utilization rates between men and women, but exhibit much sharper socio-economic disparities \cite{duggan2015mobile}. 41\% of Internet users with an annual income over \$75,000 use LinkedIn, while only 21\% of the same income bracket use Twitter. In terms of race, 47\% of African-American Internet users utilize Instagram compared to 21\% of white Internet users; this dichotomy is switched on Pinterest, where 32\% of white and 23\% of African-American Internet users maintain accounts.

These population biases like age, gender, ethnicity, and socio-economic status are potentially critical to researchers in election prediction and public health. Indeed,~\cite{gayo2011don}, appealing to the ``Literary Digest'' poll of 1936, highlights the dangers of heavily skewed sample populations when attempting to make statements about election outcomes when the demographics of those who use SM may differ significantly from the demographics of those who are likely to vote (e.g., seniors, higher educational attainment). Some election studies have taken note of this:~\cite{tumasjan2010predicting} observed that although sampled demographics were heavily skewed, the authors make no attempt to correct this bias and instead suggest the sample may be a good representative population of ``opinion leaders''. The work of ~\cite{sang2012predicting} attempts to rectify known population skew using a weighting scheme to more accurately reflect the true electorate population. Many studies~\cite{shi2012predicting, skoric2012tweets, gaurav2013leveraging}, however, make no mention of this potential bias.

Sampling strategies themselves can also be problematic. For research involving voluntary data-sharing (e.g., ~\cite{de2014characterizing,de2013social}), recent work has shown that self-reported Internet use is generally unreliable~\cite{scharkow2016accuracy}. Additionally,~\cite{salathe2010high} show that not only do SM users' network of real-world interactions differ from their SM interactions, this difference is typically larger than that perceived by the user. Keyword search methods of data sampling are subject to numerous linguistic factors that lead to bias \cite{vaughan2004search, tufekci2014big}. The work of \cite{song2016volunteerism} for example, attempts to predict a LinkedIn user's willingness to volunteer with some apparent success, but their model is trained on users who include ``volunteer'' in their profile in addition to all of those user's LinkedIn connections. The model is biased toward identifying not just volunteers, but volunteers who choose to self-identify themselves as such. 
When attempting to construct friend-follower graphs, connectivity-based sampling methods lead to unrepresentative sample populations while random population sampling leads to erroneous connectivity traits \cite{leskovec2006sampling}. 

And biases inherent to the data analyzed are not the only danger for SM researchers. Creating a predictive model entails learning a relationship between the training data (from SM) and some future real-world attributes. The advantage of SM data is often its large quantity, which lends itself well to statistical modeling. At the same time, inappropriate use of statistical techniques can easily lead to a problem known as \emph{overfitting} \cite{hawkins2004problem}. Overfitting occurs when a statistical model does not just learn the target relationship, but also captures the peculiarities and randomness inherent to the data. An overfit model may perform well under k-fold cross validation, but real accuracy will suffer when confronted with true hold-out validation data. For example, Google FluTrends showed incredible promise in the early stages of research \cite{cook2011assessing}. However, the model exhibited classic symptoms of overfitting, where model performance suffered greatly once confronted with new data \cite{olson2013reassessing}.

Because an overfit model cannot necessarily be detected based solely on its results, researchers must preemptively take measures to ensure their models will be able to generalize to novel data. For instance, in predicting depression in SM users~\cite{de2013predicting,de2013social} apply principle component analysis (PCA) to determine relevant data features, which helps in preventing overfitting by eliminating feature redundancy in the data. Overfitting can also be combated through the use of certain model training schemes, as in~\cite{lampos2013user} (via special choice of regularizers for outlier data) and discussed by~\cite{sarle1996stopped,srivastava2014dropout} (via dimension reduction).

These concerns, however, are often only paid lip service and precautions are taken infrequently \cite{gaurav2013leveraging, ruths2014social}. For example, in \cite{tumasjan2010predicting}, the authors indicate that 4\% of users were responsible for 40\% of the sampled data but no caution was expressed for the possibility that the model might overfit, learning predictions based on a small handful of users who may not generalize well to other elections. 
In the case of Twitter, given that many studies search for specific keywords or users with specific characteristics, features may be very highly correlated. In these cases, training data itself can invite bias if the models that utilize them are reinforced by too much repetition. Because SM modeling is almost entirely data-driven, overfitting should be a constant concern.

Although general SM data biases are a consistent issue in the field, there are a number of possible methods researchers can use to ensure the robustness of their results. Demographic biases should be quantified and accounted for whenever possible.  Filtering based on keywords and user characteristics are easy techniques to reduce data noise but introduce biases and should be used sparingly, replaced by alternative noise reduction methods discussed in the previous section. Finally, model overfitting should be recognized as a serious possibility and avoided through the use of certain model training schemes. Accounting for these biases not only leads to a better understanding of underlying processes in SM, but also helps ensure models can be applied beyond the data they were trained on.

\subsubsection*{Improving Generalizability}

One of the biggest concerns for SM data biases is that results which are successful in one context may fail in others. A good forecasting model should make accurate predictions not just on data from today, but also on data from tomorrow. If a forecasting model can only be applied to data from one particular location in one particular year, the model lacks the ability to forecast in a useful fashion. To accomplish this, the model needs to have learned a relationship between SM and real-world events which are unlikely to change over time. In essence, a model needs to be robust enough to generalize to novel data, not knowing ahead of time what that new data might look like. Failure to generate robust predictions might lead to poor performance whenever major changes in the world occur, often exactly when accurate predictions are most valuable. The ability to generalize findings should thus be of primary interest to researchers attempting to forecast the future, and while data biases and overfitting are two factors which can lead to generalization issues they are far from the only concerns.

The most basic cause for model generalization problems comes from a mismatch between the data a model is trained with and the data which will eventually be used in forecasting. As noted previously, it is not always possible to know what this mismatch might entail, but there are a variety of common issues that researchers can attempt to address. In particular, we note cases where poor performance is obtained due to usage of data which is too narrow in scope, where models are not evaluated across multiple possible domains of interest, and where learning occurs with data that comes from a narrow window in time.

Learning from a very narrow set of data is a large problem within the field of SM research. By narrow, we mean simply that the data may be insufficiently diverse to allow a model to perform well at a wide range of conditions. For instance, consider the case of nowcasting user political preferences on Twitter. If the goal is to detect preferences for all users, then the methodology of~\cite{volkova2015online} would be inappropriate. In their study, they trained using a small set of users who had self-identified as either Democrats or Republicans, which may not be representative of general users on Twitter. Similarly, the methods of~\cite{de2016social} in forecasting protest volume rely on a set of hashtags which were identified after the fact by mining Wikipedia pages. Applying this work to future events would require additional methods for automatically identifying the sets of hashtags relevant to an event, but~\cite{tufekci2014big} warn that any analysis based on filtering for hashtags or keywords is likely to produce a dataset with particularities specific to those keywords which may not be generalizable to future events. 

The possibility of very fragile models based on keywords alone is demonstrated in~\cite{tumasjan2010predicting}. They report that the simple percentage of mentions on Twitter for political parties in Germany reflects the share of the vote each party will win. While a very exciting premise, as~\cite{jungherr2012pirate} point out, the results are not generalizable and rely critically on excluding the German ``Pirate Party'', which was mentioned on Twitter that year more than any other political party yet garnered only 2.1\% of the vote. Changing the days of data collection likewise had major effects on the election forecast, indicating poor robustness.

One way for researchers to overcome these issues of narrow data involves learning from data over multiple SM platforms. This strategy has been used very successfully in the area of user demographic nowcasting, where relationships between user demographics and SM behavior might vary from platform to platform. For instance, \cite{sap2014developing} infer a user's age and gender based on the words they use, learning from Twitter, Facebook, or blogs either separately or together. A model trained on Facebook alone will perform well on data from Facebook, but does much more poorly on data from other sources. Models trained on a variety of sources may perform more poorly on data from any individual SM platform, but the results are more robust to changes in data source. In the work of  \cite{song2016volunteerism,jia2016fusing} an argument is made that integrating information from multiple SM platforms increases robustness of results and further has the benefit that information from different platforms is often complementary. User profiles on LinkedIn, for instance, generally contain information on educational achievement while Facebook users are much more likely to list their gender~\cite{song2016volunteerism}.

Text-based geolocation from data on a variety of SM platforms and languages is likewise explored by \cite{wing2014hierarchical}. Because their corpora are not equatable, they cannot learn over all the data at once, but their results demonstrate the utility of including more than one evaluation. They posit two models, which perform roughly equally well on data from Twitter. If they had only examined Twitter data, as in most studies on SM forecasting, they might reasonably have chosen the simpler of the two models. Unfortunately, this simpler variant performs much worse on data from Wikipedia and Flickr, a finding which would not have been recognized without investigating multiple data sources. A similar case of evaluating on multiple data sources can be found in detecting and forecasting depression~\cite{de2013predicting,de2013postpartum,de2014characterizing,tsugawa2015recognizing,de2016discovering}. While these studies differ slightly in their methods, they make use of the same basic text features~\cite{pennebaker2001linguistic}. The methodology has found positive results using Twitter~\cite{de2013predicting,de2013postpartum,tsugawa2015recognizing}, Facebook~\cite{de2014characterizing}, and Reddit~\cite{de2016discovering} and while most work has been on English-speaking SM users,~\cite{tsugawa2015recognizing} validate the findings for Japanese-speaking users as well. The method has found success in detecting depression~\cite{de2013predicting}, but also for forecasting postpartum depression~\cite{de2013postpartum,de2014characterizing} and forecasting whether a depressed individual will start thinking about suicide~\cite{de2016discovering}. By validating the methodology on such a wide variety of data sources, researchers can feel more confident that the same methods will be useful when applied to similar prediction tasks.

Incorporating non-SM data into forecasting models can also be a useful tool for increasing model robustness. By linking the model to data which is often less noisy and whose relation to prediction is better understood, studies which incorporate non-SM data are often able to achieve better forecasting performance. This practice is more common in some areas than in others. For instance, to detect adverse drug effects researchers make use of extensive domain-specific knowledge about drug side effects from non-SM databases~\cite{segura2014detecting,bian2012towards,feldman2015utilizing,yates2015extracting}. Incorporating non-SM data in this case makes side effect identification much simpler, but does not preclude models from detecting side effects which were previously unknown~\cite{feldman2015utilizing,yates2015extracting}, an important step for public health researchers.
Similarly, non-SM data can be used to enhance demographic prediction by matching unknown users against known demographic patterns~\cite{culotta2014estimating,culotta2015predicting}, geolocation by modeling population-level traffic patterns~\cite{wang2015regularity}, or by mapping place names to locations~\cite{schulz2013multi} and election prediction by factoring in variables known to affect election outcomes~\cite{williams2008political}. 

Lastly, in many cases it is important that forecasting models be trained on data that is representative of the timescale being used for prediction. For instance,~\cite{lampos2012nowcasting} use tweets from five cities in the UK to predict rainfall. Because they only have one year's worth of data, the model is always trained on 10 months and asked to predict rainfall for the remaining two. Even with such limited training data, the model does quite well, with the notable exception of predicting the weather in July, which is both a summer month, with individuals tweeting about sunny, outdoor activities, but was also the second most rainy month in the dataset. If training had utilized a longer timespan of data, such regular annual patterns could be forecast more easily. The same can be said of election outcome predictions, where training typically occurs for a single set of elections~\cite{chen2012twitter,razzaq2014prediction,shi2012predicting,cameron2015can,sang2012predicting,tumasjan2010predicting,skoric2012tweets}, which may or may not be representative of other elections past and future. Studies making use of data from multiple elections will be necessary in order to make statements regarding election forecasting that might be more reliable.

In summary, researchers should constantly be aware that decisions they make regarding training data and evaluation are likely to have a considerable impact on their ability to forecast for particular use cases. We advise researchers to make use of their domain expertise in order to determine what aspects of their model are most in need of generalization. Training using data from multiple SM platforms may be wise for a task such as protest forecasting where protesters may in future adopt a platform other than Twitter, but may be unwise for a task where only a single platform is required and is unlikely to be replaced. Training using data over an extended period of time may be appropriate when temporal patterns are relevant as with rainfall, but may be inappropriate if historical patterns have changed so dramatically as to be irrelevant. Although there is no one solution to problems of model generalization, future researchers would do well to consider how the general guidelines presented here might apply for their own use cases.

\subsubsection*{Incorporating Domain-Specific Knowledge}

Research into SM forecasting has largely found success thanks to robust statistical models which take advantage of large quantities of SM data~\cite{gayo2013power}. As we mentioned previously, applying canonical machine learning models can help researchers overcome the tremendous noise in SM data, but leaves open the possibility of overfitting, especially when data of only one type is used. Careful choice of data sources, model techniques to account for biases, and evaluation on multiple data sets can all help to overcome some of the limitations to these types of models. An additional avenue which has seen great success in many areas of SM prediction is the use of domain-specific knowledge in order to augment statistical models.

By domain-specific knowledge, we mean here the knowledge and theory specific to a particular task or field which has been validated by existing research. By incorporating patterns that are already known, researchers can point their statistical models in the right direction.  Not only can this improve model results, but it can also help to ensure generalizability. For instance, consider the detection and forecasting of depressive symptoms in SM users. A great deal is known from psychology about various types of depression and this knowledge has largely been incorporated into existing predictive research~\cite{de2013predicting,de2013postpartum,de2014characterizing,de2016discovering,tsugawa2015recognizing}. In addition, the choice of ground-truth data---study volunteers and random control data in the case of \cite{de2013predicting, de2013postpartum, de2014characterizing}---and how to manage the task of validation---confirmation bias and self-reporting---is well understood. Although all of this work relies on relatively naive text analysis~\cite{pennebaker2001linguistic}, research has been generalized well across datasets. Knowledge from psychology identifies the underlying behaviors which are linked to depression and which are expressed in SM usage. Because researchers are able to take advantage of this knowledge, they can achieve reasonable accuracy even with naive methods that do not take advantage of the scale of SM data.

In the case of depression, domain-specific knowledge manifests itself in terms of choosing model features which are linked to the target prediction of interest. Domain knowledge can also be incorporated into the structure of the forecasting model itself. Much work in geolocation now- and forecasting is built upon the knowledge that individuals tend to revisit locations they have been to before~\cite{wang2015regularity}. This fails, however, to take advantage of the patterns between individuals which exist. Results for SM users with little historical data can be improved by assuming they are similar to the general population~\cite{zhang2016eigentransitions}. The best results in the field come from models which specifically attempt to model what is theoretically known as \emph{homophily}, individuals who associate with one another are more likely to share travel patterns~\cite{backstrom2010find,schulz2013multi,zhang2016gmove}. Finding ways of incorporating sociological knowledge has allowed researchers to greatly improve location forecasts in spite of the fact that most location information in SM is quite sparse~\cite{chen2013interest} and the fact that physical, daily interaction networks and virtual SM interaction networks differ greatly from one another \cite{eagle2006reality, salathe2010high}.

Work in demographics prediction from~\cite{volkova2014inferring,Abbasi:2014:SLU:2631775.2631796} likewise use a theory-driven approach, using social influence theory in order to construct scalable network features for accurately detecting user preferences on Twitter. Indeed, the authors note that many classical machine learning models, such as logistic regression, fail to capture interrelations between users, and instead represent each user or tweet as an independent instance within the data. However, this clearly glosses over the network structure between Twitter users following and mentioning one another. The authors leverage this domain knowledge by assuming that users that follow one another likely have similar interests, and more specifically that influential users can be used to identify sub-groups of users that likely have similar interests through following or other interactions.

Failing to incorporate domain-specific knowledge or capture known or even hypothesized dynamics within the physical system of interest has been blamed for a growing number of bad outcomes in SM prediction research. Likely the most noteworthy example is Google Flu Trends \cite{lazer2014parable}. Although using relative search volume for symptoms of influenza and influenza-like-illness seems logical, there is little to no theoretical basis for this relationship. Indeed, following the same logic in terms of search volume, stock market prediction faced the same problem when \cite{ball2013counting} showed that the words ``colour'' and ``restaurant'' were the second and third best search term predictors of stock market movements. The authors had no theoretical basis on which to conclude that these results could be valid beyond what naive data analysis had shown them. Even the best predictor, ``debt'', was not entirely clear to the authors as to why it performed so well.

\section*{Conclusion}

In this systematic literature review we examine the ability of SM data to forecast real world events and characteristics across a variety of disciplines. We have focused our review toward answering two questions: Can SM be used to predict the future, and if so, how is this best accomplished? 

First, the good news: in addressing our first research question, we find that SM data has been used to make accurate forecasts across all of the disciplines examined. Additionally, topics that can be shown to be directly relevant to SM users and how they interact with SM make more successful predictions, such as user location, user demographics, and civil unrest. The bad news: in addressing our second research question, we detail four major pitfalls which have made SM prediction difficult. Noisy data, SM data biases, lack of generalizability, and difficulty incorporating domain-specific knowledge and theory lead to a fundamentally complex prediction task.

For each of these pitfalls, we examined the literature to find papers which best overcame these difficulties identifying best practices. These include, but are not limited to 1) carefully filtering out irrelevant information, such as by learning appropriate keywords \cite{baldwin2013noisy}, 2) incorporating known SM data biases by, for example, factoring in the effect of skewed demographics, 3) avoiding overfitting models to ensure predictions will be robust to future data by only incorporating relevant data features during model training such as in \cite{de2014characterizing},  as in \cite{sang2012predicting}, and 4) appealing to domain knowledge and theory, potentially through validation studies like \cite{salathe2010high}. By following these best practices, future researchers will better be able to make use of SM data, avoiding mistakes in past research which have led to poor performance.

\bibliography{litreview}

\end{document}